\pdfoutput=1
\documentclass[12pt,preprint]{aastex}

\usepackage{color}
\usepackage{ulem}
\usepackage{graphicx}
\usepackage{lscape}

\slugcomment{Accepted for Publication, AJ, Oct. 2015. Preprint}

\shorttitle{Young Stars in DR15}
\shortauthors{Rivera-G\'alvez et al.}

\begin{document}

%% LaTeX will automatically break titles if they run longer than
%% one line. However, you may use \\ to force a line break if
%% you desire.

\title{The Young Stellar Population of the Cygnus-X DR15 Region}

%% Use \author, \affil, and the \and command to format
%% author and affiliation information.
%% Note that \email has replaced the old \authoremail command
%% from AASTeX v4.0. You can use \email to mark an email address
%% anywhere in the paper, not just in the front matter.
%% As in the title, use \\ to force line breaks.

\author{S. Rivera-G\'alvez\altaffilmark{1}, C. G. Rom\'an-Z\'u\~niga\altaffilmark{2,5},
E. Jim\'enez-Bail\'on\altaffilmark{2}, J. E. Ybarra\altaffilmark{2}, J. F. Alves\altaffilmark{3,5}
and Elizabeth A. Lada\altaffilmark{4}}

%% Notice that each of these authors has alternate affiliations, which
%% are identified by the \altaffilmark after each name.  Specify alternate
%% affiliation information with \altaffiltext, with one command per each
%% affiliation.

\altaffiltext{1}{Facultad de Ciencias, Universidad Aut\'onoma de \\
Baja California, Ensenada, M\'exico}
\altaffiltext{2}{Instituto de Astronom\'ia, Unidad Acad\'emica en Ensenada,\\
Universidad Nacional Aut\'onoma de M\'exico, Ensenada 22860, M\'exico}
\altaffiltext{3}{Astronomy Department, University of Vienna, Austria}
\altaffiltext{4}{Astronomy Department, University of Florida, USA}
\altaffiltext{5}{Formerly at Centro Astron\'omico Hispano Alem\'an, Glorieta de la Astronom\'ia, S/N, Granada, Spain}

%% Mark off your abstract in the ``abstract'' environment. In the manuscript
%% style, abstract will output a Received/Accepted line after the
%% title and affiliation information. No date will appear since the author
%% does not have this information. The dates will be filled in by the
%% editorial office after submission.

\begin{abstract}
We present a multi-wavelength study of the young stellar population in the Cygnus-X DR15 region. We
studied young stars forming or recently formed at and around the tip of a prominent molecular pillar and an infrared
dark cloud. Using a combination of ground based near-infrared, space based infrared and X-ray data, we
constructed a point source catalog from which we identified 226 young stellar sources, which we
classified into evolutionary classes. We studied their spatial distribution across the molecular gas structures
and identified several groups possibly belonging to distinct young star clusters. We obtained samples
of these groups and constructed K-band luminosity functions that we compared with those of artificial
clusters, allowing us to make first order estimates of the mean ages and age spreads of the groups. We
used a $^{13}$CO(1-0) map to investigate the gas kinematics at the prominent gaseous envelope of
the central cluster in DR15, and we infer that the removal of this envelope is relatively slow compared
to other cluster regions, in which gas dispersal timescale could be similar or shorter than the circumstellar disk dissipation
timescale. The presence of other groups with slightly older ages, associated with much less prominent
gaseous structures may imply that the evolution of young clusters in this part of the complex proceeds
in periods that last 3 to 5 Myr, perhaps after a slow dissipation of their dense molecular cloud birthplaces. 

\end{abstract}

%% Keywords should appear after the \end{abstract} command. The uncommented
%% example has been keyed in ApJ style. See the instructions to authors
%% for the journal to which you are submitting your paper to determine
%% what keyword punctuation is appropriate.


%% From the front matter, we move on to the body of the paper.
%% In the first two sections, notice the use of the natbib \citep
%% and \citet commands to identify citations.  The citations are
%% tied to the reference list via symbolic KEYs. The KEY corresponds
%% to the KEY in the \bibitem in the reference list below. We have
%% chosen the first three characters of the first author's name plus
%% the last two numeral of the year of publication as our KEY for
%% each reference.

%% Authors who wish to have the most important objects in their paper
%% linked in the electronic edition to a data center may do so by tagging
%% their objects with \objectname{} or \object{}.  Each macro takes the
%% object name as its required argument. The optional, square-bracket 
%% argument should be used in cases where the data center identification
%% differs from what is to be printed in the paper.  The text appearing 
%% in curly braces is what will appear in print in the published paper. 
%% If the object name is recognized by the data centers, it will be linked
%% in the electronic edition to the object data available at the data centers  
%%
%% Note that for sources with brackets in their names, e.g. [WEG2004] 14h-090,
%% the brackets must be escaped with backslashes when used in the first
%% square-bracket argument, for instance, \object[\[WEG2004\] 14h-090]{90}).
%%  Otherwise, LaTeX will issue an error. 

\section{Introduction \label{intro}}

\par The star forming complex of Cygnus-X region is one of the most prominent features in our Galaxy. Originally detected as an extended region with a thermal spectrum, \citet{Piddington:1952aa} named the region Cygnus-X, in order to distinguish it from the other nearby known radio source, Cygnus-A. Cygnus-X is composed of several OB associations, dozens of embedded stellar clusters, hundreds of HII regions and over 40 known massive protostars \citep[see][for an extensive review]{Reipurth:2008aa}. The young star population in Cygnus-X is currently interacting with one of the most massive molecular cloud complexes in the Galaxy, with a total mass of $3\times 10^6\ \mathrm{M}_\odot$ \citep{Schneider:2006aa}, as well as the X-ray emitting Cygnus Superbubble. It has been proposed that Cygnus-X could be the precursor of a globular cluster \citep{Knodlseder:2000aa}.

\par To study the interaction between recently formed star clusters and their surrounding medium in Cygnus-X could provide very important clues about the present evolution of the complex. It is particularly important to focus on the numerous embedded cluster populations and to compare the properties of clusters across the region. For instance: how embedded clusters proceed from formation to emergence from their parental gas clumps in such a strong ionizing medium? Is cluster formation or evolution in Cygnus-X determined by the local environment? What are the time scales from formation to gas dispersal? Also, what happens after cluster emergence: are subsidiary clusters destined to disperse or could they end up swelling the ranks of the main association? In any case, we should expect that multiwavelength analysis of the young star clusters in Cygnus-X will help to fine tune current ideas about the formation and evolution of embedded stellar clusters or groups.

\par
\begin{figure*}
\begin{centering}

\includegraphics[width=6.0in]{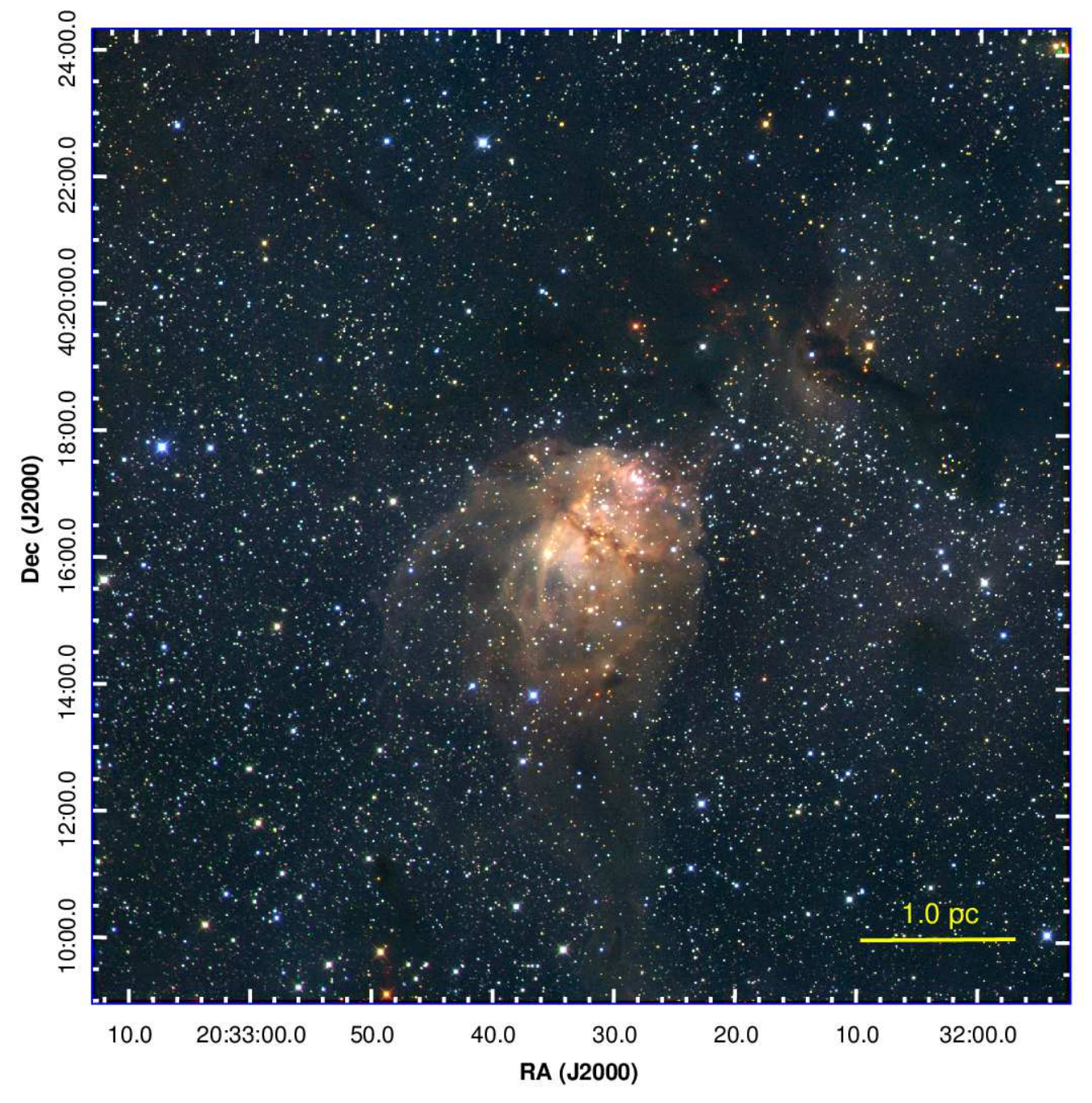}
\caption{False RGB (KHJ) representation of the cluster DR15 in near-IR wavelengths. Images are log scaled. The DR15 cluster is located near the center of the image, surrounded by a conspicuous nebulous envelope, resulting from reflections in photodissociation regions and expanding parental gas near and around the young sources. The two dark lanes populated with red sources at the top of the image are part of the active infrared dark cloud.}
\label{fig:RGB}
\end{centering}

\end{figure*}

\begin{figure*}
\begin{centering}

\includegraphics[width=6.0in]{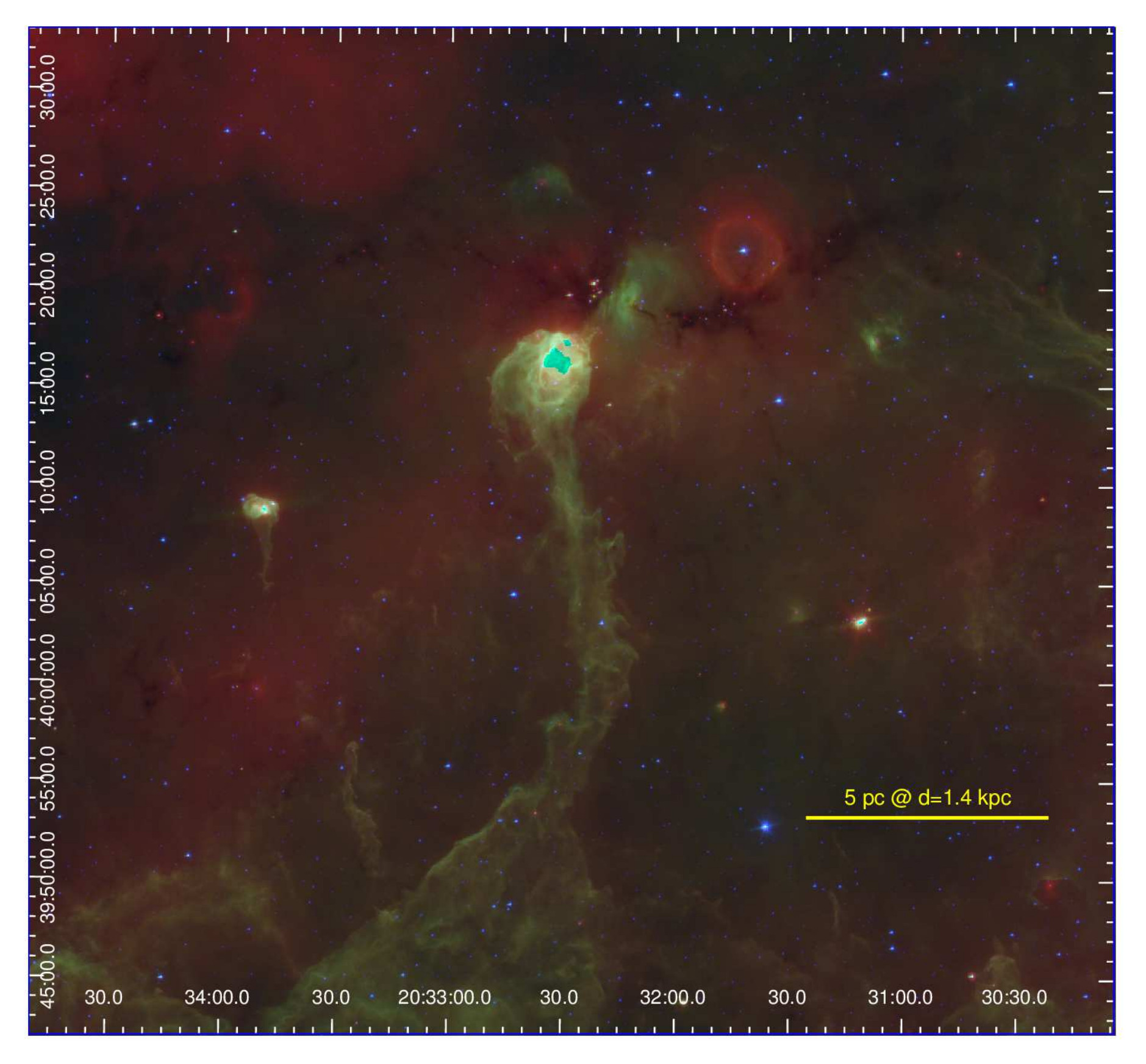}
\caption{False RGB (Spitzer MIPS 24 $\mu$m, IRAC [8.0], IRAC [4.5]) image of the DR15 region. Images are log-scaled. The DR15 pillar is located at the center. Bright PAH emission and reflections saturated the pixels at the center of the DR15 cluster shell in the IRAC [8.0] and MIPS 24 $\mu$m channels, thus the cyan-colored region. Other fainter, molecular pillar structures are visible at both flanks of the main pillar. The dark infrared cloud at the north shows recent star formation activity and extends westward towards the LBV source  G79.29+0.46, located at the center of a bright red shell.}
\label{fig:spitzer}
\end{centering}
\end{figure*}

 In this paper we selected to study one of the most prominent embedded cluster populations in Cygnus-X: the region DR15, also listed as cluster 10 in the survey of \citet{Dutra:2001aa}. The DR15 region has been related to the HII region G79.306+0.282, the source IRAS 20306+4005/FIR-1 \citep{Campbell:1982aa}, and sources IRS 1, 2 and 3 in the list of \citet{Kleinmann:1979aa} \citep[see also][]{Odenwald:1990aa}. DR15 is located in the Cygnus-X South region, located at an estimated distance of 1.4 kpc \citep{Rygl:2012aa}. The cluster contains one prominent far-infrared source (FIR-1), which marks the location of a compact HII region formed by a pair of B type embedded sources \citep{Odenwald:1990aa,Oka:2001aa,Kurtz:1994aa}, however the cluster hosts various intermediate to massive stars which create a rather complex structure of photodissociation regions, forming a nebulous envelope which glows brightly in infrared images (see Figure \ref{fig:RGB}). DR15 sits at the tip of a long (about 10 pc), filamentary pillar that extends into the southern edge of the central OB from DR12, lying in projection about 1 degree south from the Cyg-X3 star. The structure of the pillar appears to be protruding from the DR12 ridge (see Figure \ref{fig:spitzer}). The DR12-15 region in Cygnus-X possibly lies in front of the OB2 association, along our line of sight. A large filamentary infrared dark cloud (IRDC) with active star formation lies to the north and west of the cluster. This IRDC appears to be kinematically independent from DR15, as shown by \citeauthor{Schneider:2006aa} analysis. Recently, the western segment of the IRDC, has been shown to host a young stellar cluster in interaction with the Luminous Blue Variable (LBV) source G79.29+0.46 \citep{Rizzo:2008aa,Jimenez:2010aa}. Distance to G79.29+0.46 has been also estimated to be 1.4 kpc \citep{Rizzo:2014aa,Palau:2014aa}.   

\par Our main goal in this study is to track the progression of the star formation in the DR15 cluster and its immediate surroundings, by looking at the properties and distributions of the young stellar population. DR15 is especially interesting for being at a very specific stage at which it is already emerging from its parental cloud (some members and parts of the reflection nebulosities are already detectable in visible wavelengths). Moreover, the interesting case it presents at being apparently formed at the tip of a filamentary structure and next to a much younger star forming spot in the IRDC, is worth of detailed attention. 

\par The paper is organized as follows: in Section \ref{obs} we describe the datasets used in this work. Section \ref{analysis} is dedicated to describe the procedures we followed to acomplish a description of the history of star formation in the region, a result which we discuss in Section \ref{discussion}. Finally,  a brief summary of our results can be found in Section \ref{discussion}.

\section{Observations}
\label{obs}

For the first part of our analysis, we combined high-quality deep near-IR observations with images and catalogs from the Spitzer Cygnus-X Legacy Survey \citep[][hereafter CXLS]{Hora:2009aa}. We also made use of archival data from the Chandra X-Ray Observatory. By combining these datasets we prepared a multi-wavelength photometry catalog for the stellar population in DR15, which we use to reconstruct the history of star formation in the region.

\subsection{Near Infrared Observations}
\label{obs:nir}

\par Near-infrared images of the Cygnus-X DR15 region were obtained with the OMEGA 2000 camera at the 3.5m telescope of the Calar Alto Observatory, atop Sierra de los Filabres in Almer\'ia, Spain, during the nights of February 2nd and March 3rd, 2010, with excellent weather conditions. The dataset consists of 900 second co-added exposures in the $J$, $H$ and $K$ bands (1.209, 1.648 and 2.208 $\mu$m, respectively). The seeing values --measured directly from the average FWHM of stars in the final reduced mosaics-- were 1.17, 1.13 and 0.98$\arcsec$ in $J$, $H$ and $K$, respectively. 

\par The reduction of the images and the extraction of Point Spread Function (PSF) photometry lists for all bands were performed with custom \texttt{IRAF} pipelines, and the \texttt{SExtractor} algorithm \citep{Bertin:1996aa}, following a methodology equivalent to the one described in \citet{Roman:2010aa}. 

\par We constructed a near-IR catalog by matching individual filter photometry lists with the aid of {\tt TOPCAT} \citep{Taylor:2005uq}. We replaced saturated sources and a small number of missing detections in the Calar Alto images with values from the 2MASS Point Source Catalog, obtained at the Infrared Processing and Analysis Center (IPAC, Caltech). The 10 percent photometric depth values --indicated by the average brightness at which photometric error reaches a value of 0.1 mag-- are 21.5, 20.0 and 18.75 mag in $J$, $H$ and $K$, respectively, and are good indicators of the completeness of the data. These limits are enough to sample the young star population in DR15 down to 0.09 M$_\odot$ in regions of low to moderate extinction ($A_V<25.0$ mag).

\subsection{Spitzer Cygnus-X Legacy Survey data }
\label{obs:mir}

The \textit{Spitzer Space Telescope} has observed the DR15 cluster with the IRAC and MIPS detectors as part of the CXLS. We obtained archival enhanced product mosaics from the Spitzer Heritage Archive as well as a photometric catalog coincident with our region of interest directly from the CXLS Data Release 1 (DR1). The catalog contains calibrated magnitudes for sources detected with IRAC in its four cryogenic mission channels (3.6, 4.5, 5.8 and 8.0 $\mu$m), as well as in the 24\ $\mu$m channel of MIPS. Using \texttt{TOPCAT} and IDL routines we combined our near-IR and the CXLS DR1 catalogs into a single infrared photometry list with a total of 46983 sources. 

\subsection{Chandra ACIS Observations}
\label{obs:xray}

\par The  DR15 cluster was observed with the Imaging Array of the Chandra Advanced CCD Imaging Spectrometer (ACIS-I) on  2011 January 25 (ObsID 12390, P.I. Wright) with net exposure time of 39.875 ks. 

\par We processed the archival raw data using the routines from version 4.5 of the Chandra Interactive Analysis of Observations \texttt{CIAO} data analysis system \citep{Fruscione:2006aa}. We used the \texttt{chandra$\_$repro} reprocessing script to recalibrate our event data in order to ensure that consistent calibration updates were applied to the dataset.

\par The X-ray image processing was performed as follows: first, we created a exposure-corrected image from our data in the broad band (0.5 to 7 keV with an effective energy of 2.3 keV) using \texttt{CIAO/fluximage} routine with a bin size of 1. Second, we applied the \texttt{wavedetect} tool to our broad band image in order to identify potential X-ray sources. We used wavelet scales from 1 to 16.5 pixels in s.pdf of $\sqrt{2}$ and a source significance threshold of $1\times10^{-6}$.

\par
\begin{figure*}
\begin{centering}
\includegraphics[width=7.0in]{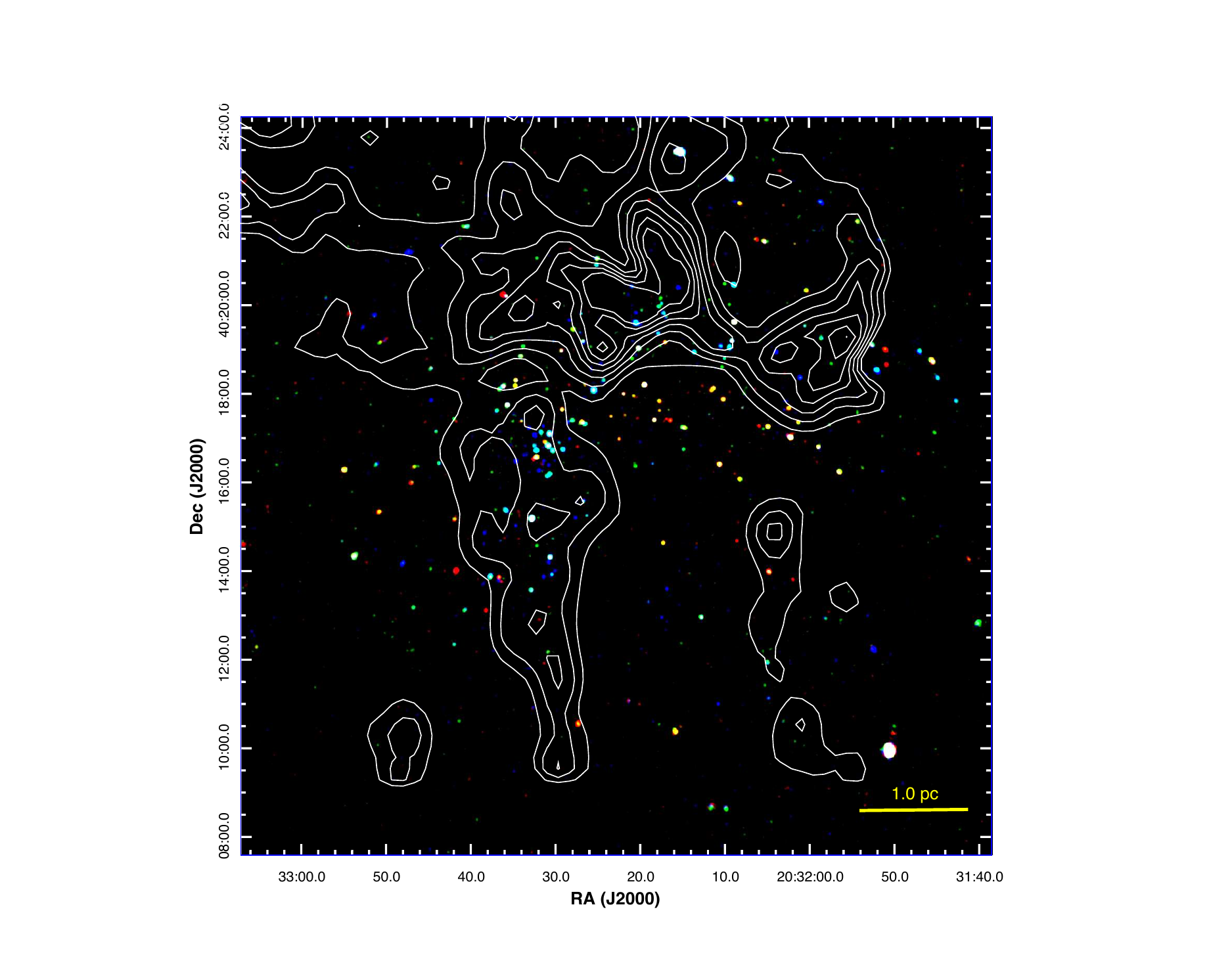}
\caption{False color image of X-ray emission in DR15, constructed from Hard (4.0-7.0 keV, blue), Medium (2.0-4.0 keV, green) and soft (0.2-2.0 keV, red) channel emission in the Chandra ACIS field. Overlaid are contours of visual extinction, $A_V$, in magnitudes from 15 to 30 in step of 2.5. See section \ref{analysis:extmapx}}
\label{fig:chandra}
\end{centering}
\end{figure*}

 We performed photometry on the list provided by \texttt{wavedetect} using the the \texttt{ACIS Extract} (AE) package \citep{Broos:2010aa}. AE permits an optimal determination of the local background and the best flux extraction apertures based on the PSF of the image. Photometry is then extracted on selected energy bands and a list of source properties, statistics and best fit spectral models is produced. We used three energy bands: Soft, from 0.2 to 2.0 keV, Medium, from 2.0 to 4.0 keV and Hard, from 4.0 to 7.0 keV. From the final list of sources produced by \texttt{ACIS Extract} we rejected those with a probability of 1$\%$ or higher of being a background fluctuation $P_B>0.01$). Our final lists contains a total of 131 X-ray sources. From these, a total of 109 (83.2\%) sources have a counterpart in our IR catalog. In Figure \ref{fig:chandra} we show a false color (RGB) map of the ACIS field for DR15 using images from the three energy bands, overlaid with contours of visual extinction from the NICEST extinction map ww constructed from near-IR data. In there, we see how sources with hard X-ray emission are preferently located in regions of high column density. This is because soft X-ray bands are prone to oscuration by dust. We see how most of the X-ray sources are associated with the molecular cloud and the embedded population, confirming that most of the point source X-ray emission in DR15 comes from young stars.

%\clearpage

\subsection{FCRAO Observations }
\label{obs:others}

\par We made use of the Five College Radio Observatory (FCRAO) $^{13}$CO(1-0) molecular radio emission map of the South Cygnus-X region from the study of \citet{Schneider:2011aa}. The map is a RA-Dec-radial velocity cube, from which we extracted molecular gas properties using standard tools from the \texttt{MIRIAD} \citep{Sault:1995aa} package.

\section{Analysis }
\label{analysis}

\par We limited our region of study to the area covered with OMEGA 2000, defined as $\mathrm{[RA,\ Dec]}=[307.969040,40.150383]\rightarrow[308.305620,40.405865]$. The analysis described below correspond to sources falling within that area only.

\subsection{Identification of YSOs }
\label{analysis:ysoid}

\par We identified Young Stellar Objects (YSOs) in the DR15 region, by applying color and brightness criteria to our multiwavelength catalog. For sources that were detected in the Spitzer IRAC and MIPS 24$\mu$m bands, we classified Class I/0 (embedded protostars) and Class II (Classic T Tauri) sources using the criteria by \citet{Ybarra:2013aa}, we also required that these sources had photometric uncertainty values less than 0.25 mag in each band.

\par Our color criteria \citet{Ybarra:2013aa} are essencially based in the color criteria of \citet{Gutermuth:2008aa} and \citet{Kryukova:2012aa}, but as explained by \citeauthor{Ybarra:2013aa} we add an additional [5.8]-[8.0] criteria for objects that do not have a MIPS 24 $\mu m$ detection. In addition, our use of $JHK$ photometry allows us to identify additional less bright candidates and with the use of X-ray data we are able to identify Class III sources that do not have an infrared excess.

\par Class III candidate sources were selected from a list of sources in the Chandra catalog that match in position with an infrared point source, after removing those corresponding to Class I/0 and Class II candidates. We added the requirement that $$J-H\ge0$$, because sources with $J-H<0$ are most likely either galaxies or spurious sources associated with diffraction spikes of bright stars. We further depurated this first counterpart list by keeping only those sources that had no evidence of a prominent circumstellar disk (i.e., we only selected stars with mostly photospheric emission). For this last criteria we used the $J-H$ vs $K-[4.5]$ color space: 

$$J-H>1.97(K-[4.5]).$$

and for those sources that do not have a detection in [4.5] we used:

$$J-H>1.74(H-K).$$

\par The remaining list of Chandra-NIR counterparts are sources with evidence of a disk that were not previously selected as Class 0/I or II sources. We used the additional near-IR criteria of \citeauthor{Gutermuth:2008aa} to select a few more young sources from this group. If a Chandra counterpart had photometry in the first three IRAC bands, and photometric errors less than 0.1 mag in at least [3.6] and [4.5], then it was classified as a Class 0/I candidate if:

$$[4.5]-[5.8]>0.5 \mathrm{\ and\ } [3.6]-[4.5]>0.7,$$

while for Class II candidate identification, we used:

$$0.2<[3.6]-[4.5]<0.7 \mathrm{\ and\ } 0.5<[4.5]-[5.0]<1.5.$$

\par A total of 20 sources could not be classified with these criteria if they did lack a detection in one or more bands, but if they fall to the right of the reddening band in the $J-H$ vs $K-[4.5]$ color space, they could be bonafide young sources. For 13 of these sources we determined their class (I or II) by inspecting their spectral energy distrubutions (SED), which we constructed using the SED fitting tool of \citet{Robitaille:2007aa}. Four additional sources were identified as possible AGN galaxies. The 3 remaining sources did not have enough IR bands to permit a clear classification and were discarded from the list.

\par In total, we identified in our selected region a total of 226 YSOs, distributed as follows: 26 Class I/0 candidates, 155 Class II candidates and 45 Class III candidates (11, 69 and 20 percent of the total, respectively). We list all identified YSOs in the tables of the Appendix \ref{app:ysos}.

\par We matched our Class I, Class II and Class III catalogs against the catalog of \citet{Kryukova:2014aa}, out of 23 sources coinciding with our region of study, 17 were also identified as YSO candidates in this study. The 6 remaining sources lack emission is the [4.5] band and could not be confirmed as YSOs using our criteria.

\par
\begin{figure}
\includegraphics[width=3.0in]{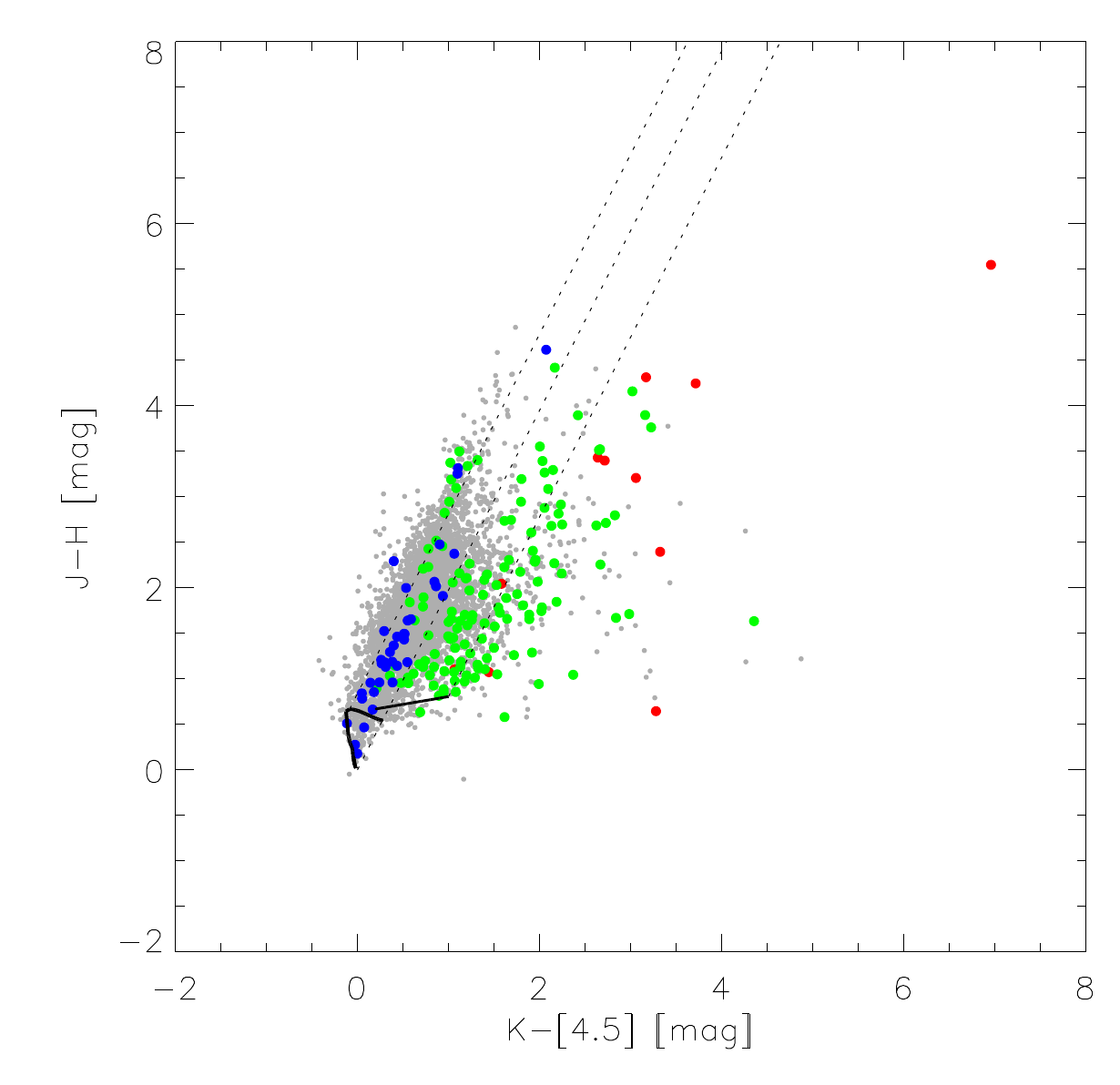}
\includegraphics[width=3.0in]{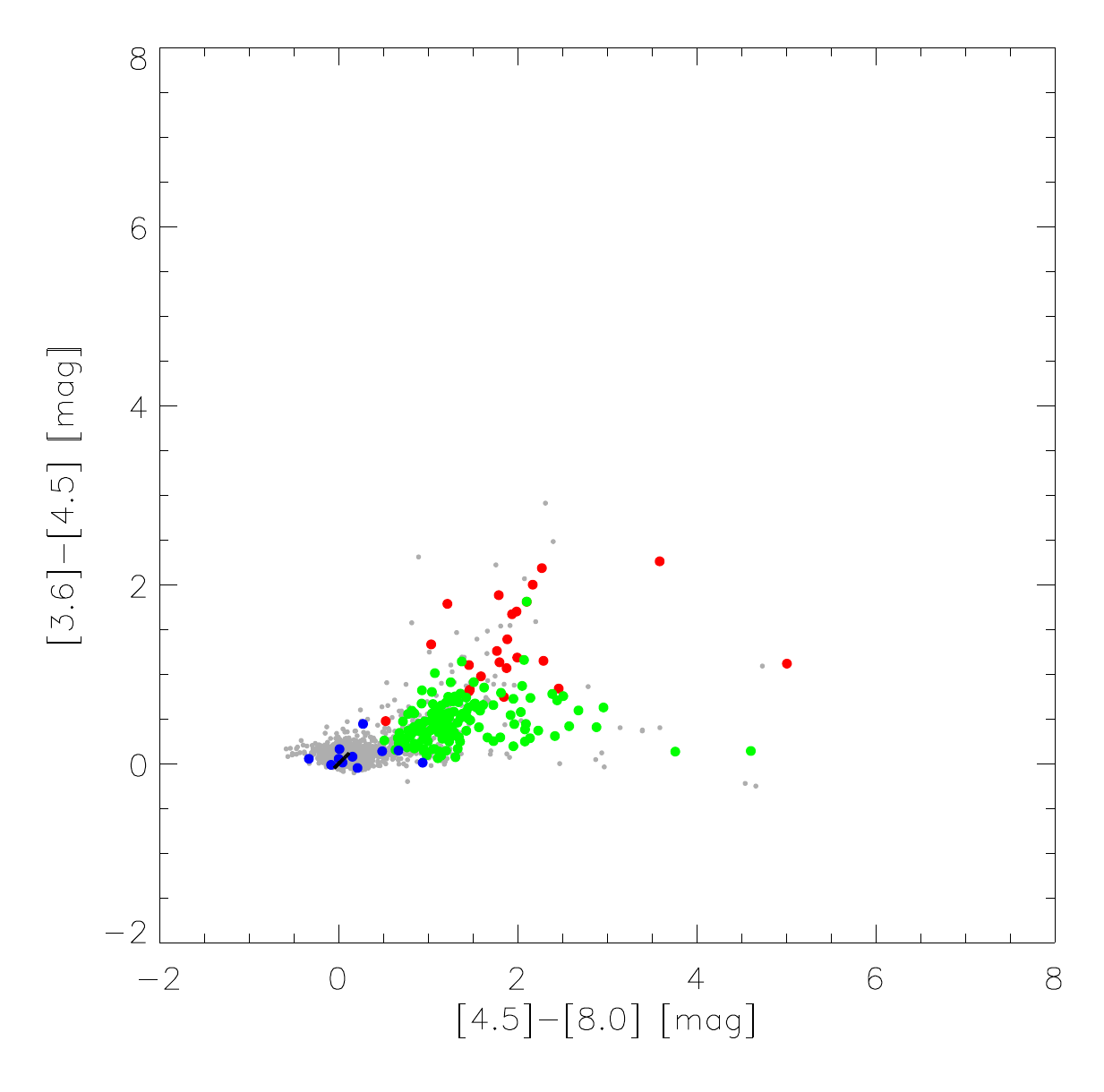}
\caption{Color-color diagrams for YSO sources in the DR15 region, Class I sources are marked with a red dot symbol; Class II sources are marked with a green dot symbol; Class III sources are indicated with a blue dot symbol. Sources marked with gray dot symbols are sources in the field without a YSO designation and photometric errors less than 0.1 mag in all bands.  a) $J-H$ vs. $K-[4.5]$ two-color diagram. The solid curve is a model for dwarf main sequence population from the Dartmouth \citep{Dotter:2008aa} grid; the solid line at its right side is the Classic T-Tauri star locus \citep{Meyer:1997aa} adapted to this color space as in \citet{Teixeira:2012aa}. b) [3.6]-[4.5] vs. [4.5]-[8.0] two-color diagram, showing a clear separation between evolutionary classes.}
\label{fig:diagrams}
\end{figure}

 In Figure \ref{fig:diagrams} we show the distribution of the classified YSOs in two different color-color diagrams. In the $J-H$ vs. $K-[4.5]$ diagram we can see that the most of Class I sources lie to the right of the reddening band indicating the presence of intense excess emission at IR wavelegths due to the stellar radiation in the dusty material of their envelopes or circumstellar disks. Class II sources also present infrared excess emission, although in a lesser way due the dust clearing within their inner disks. Then, the Class III sources lie within the reddening band or along the dwarf main sequence and lack significant infrared excess. The three groups of sources appear well separated in the [3.6]-[4.5] vs. [4.5]-[8.0] two-color diagram.

%\clearpage

\subsection{Dust Extinction and the Spatial Distribution of YSOs \label{analysis:extmap}}

\par
\begin{figure*}
\begin{centering}
\includegraphics{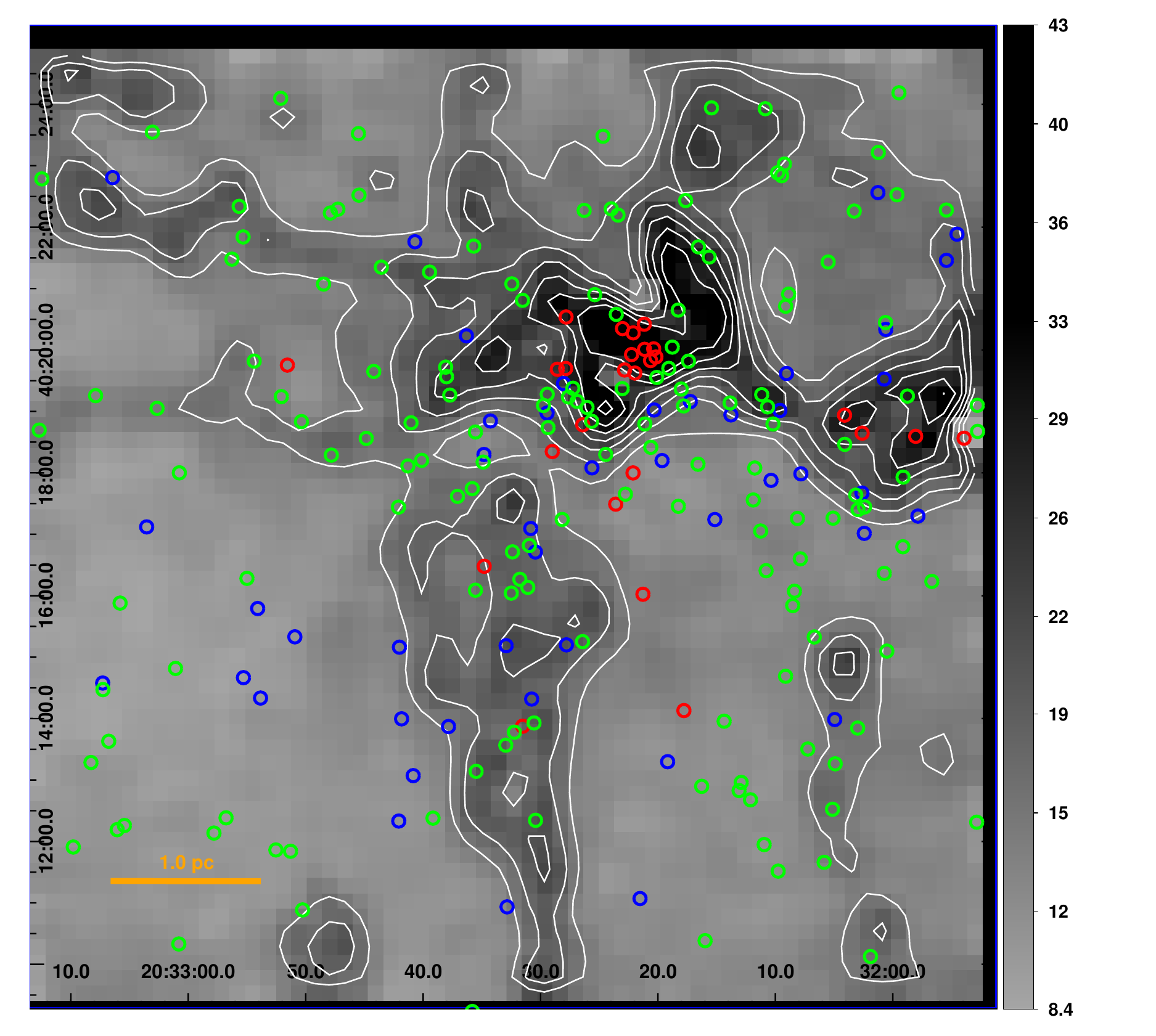}
\caption{Spatial distribution of the identified YSOs in DR15 plotted on the extinction map. The grayscale indicate visual extinction, $A_V$, in magnitudes, with contours highlighting levels from 15.0 to 30.0 in s.pdf of 2.5. The red circles indicate the Class I objects, the green circles the Class II objects and the blue circles the Class III objects.}
\label{fig:ysodist}
\end{centering}
\end{figure*}

 In Figure \ref{fig:ysodist} we plot the spatial distribution of the YSOs on a dust extinction map. This dust extinction (A$_V$) map for the DR15 region was constructed with the near-IR catalog an optimized version of the near-infrared excess (NICER) algorithm of \citet{Lombardi:2001aa}, which estimates extinction using the dust-reddened colors of stars in the background of the cloud. In order to construct the map, we removed all sources with colors indicative of intrinsic infrared excess, which would bias the estimated extinction towards higher values. We used a sigma-clipping scheme to remove outlier values from the final weighted averages at each position. The map was constructed using Nyquist sampling on a equally spaced equatorial grid, smoothing the individual extinction estimates of background stars with a Gaussian filter of 30$\arcsec$ FWHM (this implies a resolution near 0.2 pc at a distance of 1.4 kpc). In Figure \ref{fig:ysodist} we see how the map clearly resolves the filamentary structure of the IRDC in the northern part of the field and the morphology of the dust pillar on which DR15 is located. The extinction contours are limited to a maximum level of $A_V$=30 mag and further smoothed on the figure with a factor 3 boxcar, in order to remove some spurious features at the IRDC region. The resultant map shown in Figure \ref{fig:ysodist} is in good agreement with a column density map constructed from 250 to 500 micron dust emission images from Herschel/SPIRE\footnote{Herschel is an ESA space observatory with science instruments provided by European-led Principal Investigator consortia and with important participation from NASA.}, which also allows a much larger dynamic range (up to $A_V\approx$150 mag; Schneider et al., private comm.)

\par Clearly, the regions with higher column density values are those hosting a majority of the youngest stars. Using our extinction map we determined how the Class I sources are distributed in the highest density regions. Using contours of constant extinction in s.pdf of 1.0 mag, we counted the number of Class I sources above each level, and found that 23 out of 24 sources in our region of study lie above A$_V$=13.0 mag, and 20 out of 24 sources lie above $A_V$=15.0 mag. The fraction of the total of Class I sources, $N(>A_V)/N_{total}$, decreases steeply after that, with only 50 percent of the total number of sources remaining at levels above $A_V$=22.0 mag. It is also important to notice that given the filamentary morphology of the IRDC, the projected area of the map contained above each level decreases very steeply for $A_V>12.0$ mag. We also found that the surface density of Class I sources, $\Sigma_*(>A_V)$, defined as the number of sources divided by the area above a given level, deviates little from a power-law behavior with a slope $\beta \approx 2.9$ in the range $12<A_V<40.0$ mag. All of this is very consistent with a Schmidt type relation like it was found for a set of nearby Giant Molecular Clouds by \citet{Lada:2013aa}, except that in those clouds the linear regime appears to be set at a lower extinction interval. Unfortunately, our numbers are too small to attempt a Bayesian analysis like that of \citet{Lada:2013aa}, and possibly a comparison with other regions containing IRDCs would be more fair. However, the fact that we find a possible Schmidt-like behavior in a region like Cygnus-X DR15 is very interesting and worth of a dedicated comparative analysis with similar regions, which is beyond the scope of this paper.

\subsection{Properties of the youngest stars in the DR15 region \label{analysis:SED}}

\par
\begin{figure}
\includegraphics[width=2.5in]{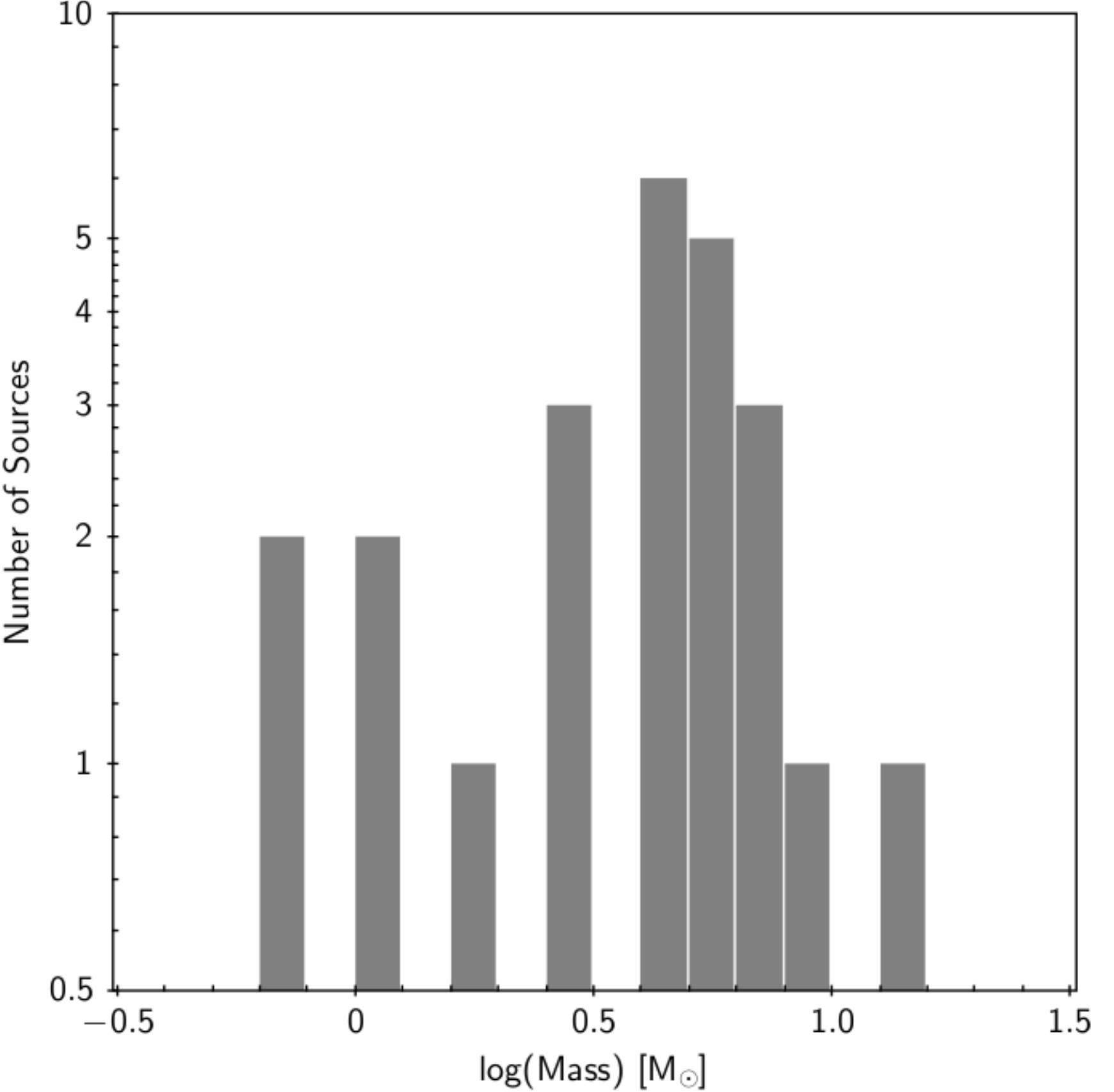}
\includegraphics[width=2.5in]{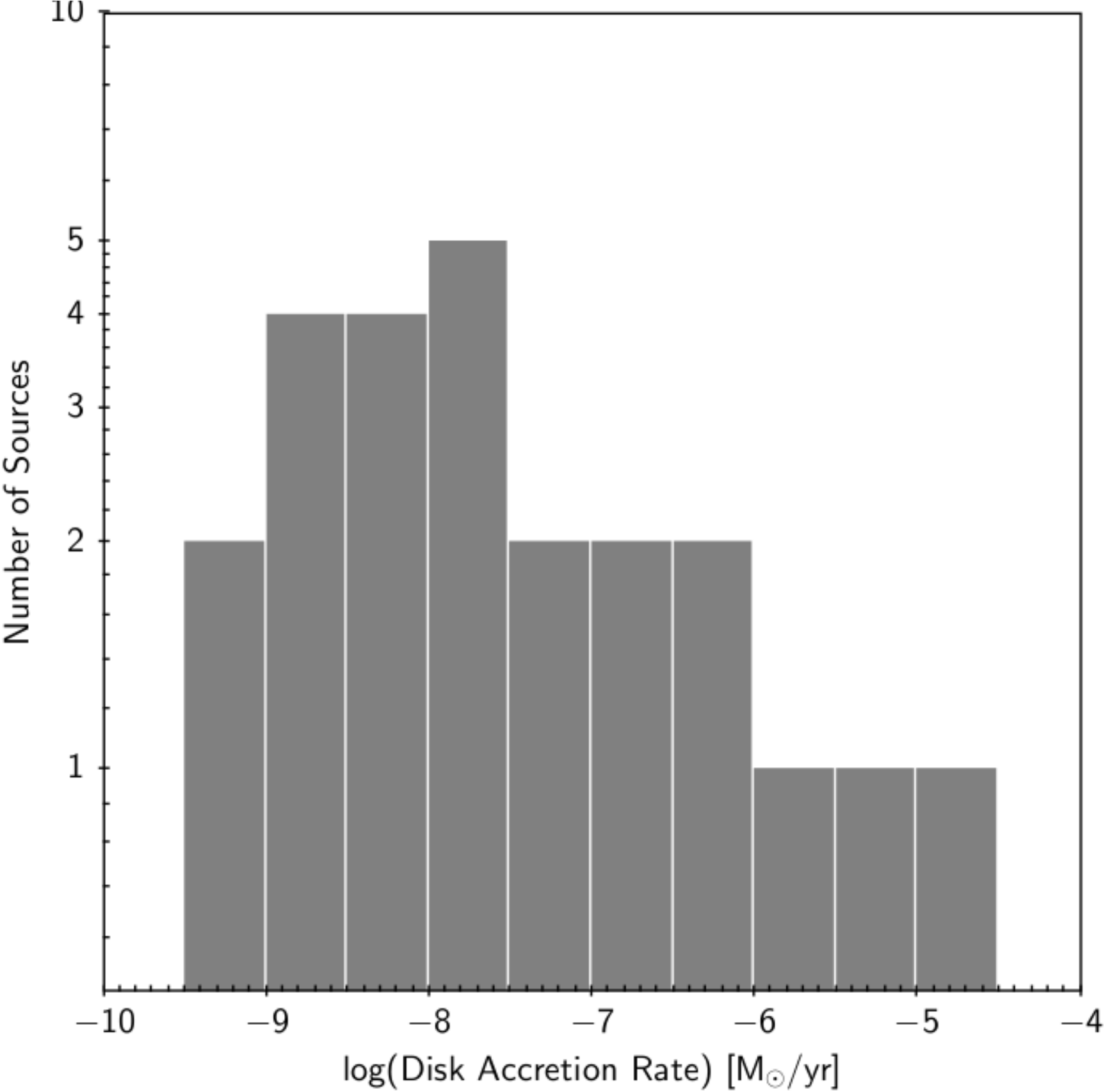}
\caption{Left: Distribution of Class 0/I candidate sources mass estimates from SED model fits. Right: Distribution of Class 0/I candidate sources disk accretion rate estimates from SED model fits.}
\label{fig:C1disks}
\end{figure}

 Using our master infrared catalogs we were able to construct spectral energy distrubutions (SED) for 24 of the Class 0/I sources we identified using the methods previously described. In most cases these sources belong to the cluster population associated with the dark infrared cloud in the northern section of our region of study. In a few cases we were able to complement these SEDs with Herschel Space Telescope PACS mid/far infrared photometry from the catalog of \citet{Ragan:2012aa} and 850 $\mu$m photometry from the catalog of \citet{Difrancesco:2008aa}. Using the SED fitting tool of \citet{Robitaille:2007aa}, we estimated some basic properties for the sources.  In Figure \ref{fig:C1disks} we show histograms of star masses and the disk accretion rates. The mass distribution suggests that the sources we are able to detect in DR15 are mostly intermediate to massive (mostly solar type and above). The accretion rates are in good agreement with estimates of typical accretion disk for sources with masses above solar \citep{Fang:2013aa}.

%\par In Table \ref{tab:SEDs} we present some of the main properties of protostars in DR15 infered from the model fits. 

\subsection{The Star Formation History of DR15 \label{analysis:sfh}}

\subsubsection{Identification of the main stellar groups in DR15}
\label{analysis:sfh:clusterid}

\par To further investigate the stellar formation history in the DR15 region we attempted to identify the individual stellar clusters present in our field of study. Two clusters are easy to identify, namely the cluster at the center of the field and the cluster embedded in the IRDC at the north. However, the map of Figure \ref{fig:ysodist} shows a good number of sources located around the molecular gas filaments, which may or may not be part of previously formed stellar clusters or groups.

\par In order to identify significant overdensities of stars in the DR15 region, we constructed surface density maps using the \texttt{Gather} algorithm of \citet{Gladwin:1999id}. This algorithm is based in turn on the \textit{nearest neighbor} method, which assigns individual surface density values to points on a two-dimensional map based on the equivalent circular area defined by the distance to a $n$th. neighbor point  (see \citet{Casertano:1985uq} for a description of this method in the particular case of stellar cluster identification; some examples of its use for embedded clusters can be found in \citet{Roman-Zuniga:2008aa,Gutermuth:2009aa} and \citet{Roman:2015aa}). The \texttt{gather} algorithm is adequate for identifying individual clusters in a relatively simple layout, optimizing the value of $n$ that defines the surface density measurement and the size of the smoothing kernel used to construct the surface density map.  

\par
\begin{centering}
\begin{figure*}
\includegraphics[width=7.0in]{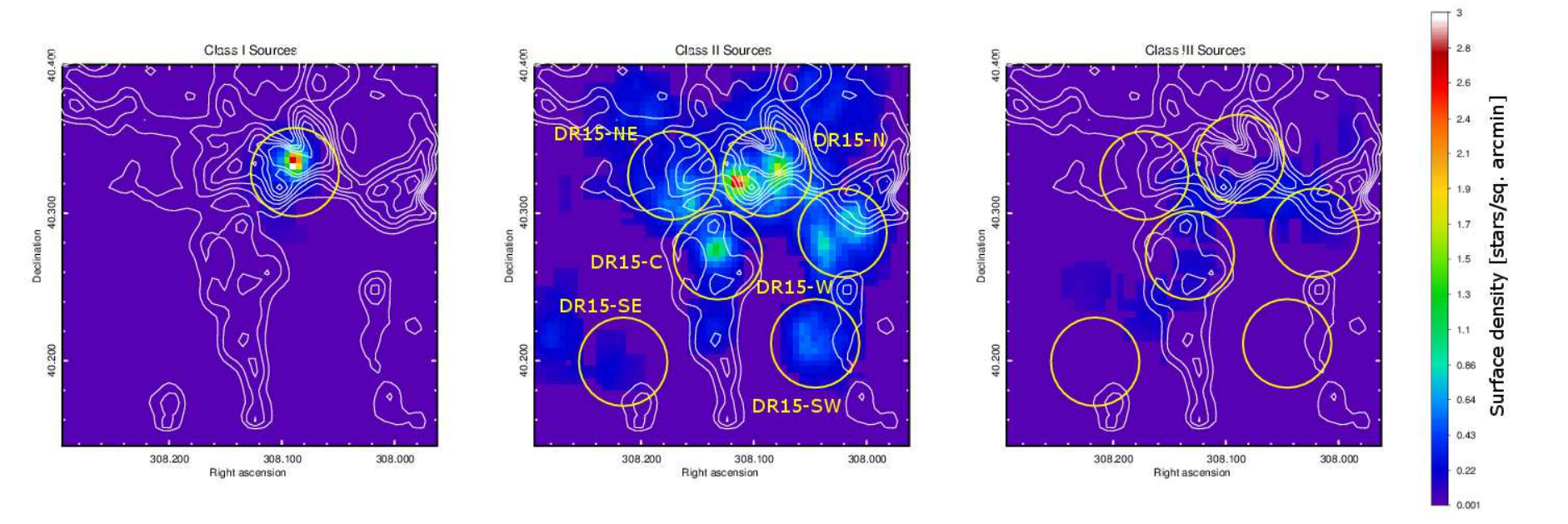}
\caption{\textit{From left to right}: Color scaled surface density maps, in units of sources per sq. arcmin, for Class I, II and III candidate YSOs in DR15, constructed with the \texttt{Gather} algorithm. Extinction contours are similar to Figures \ref{fig:chandra} and \ref{fig:ysodist}. Yellow circles indicate the location and selected sample extension for clusters and groups identified in the Class II map.}
\label{fig:gather}
\end{figure*}
%\clearpage
\end{centering}

 We made individual \texttt{gather} maps for each of the candidate YSO class lists. These maps are shown in Figure \ref{fig:gather}. The maps show the concentration of Class I sources at the north IRDC region. Class II sources are distributed over the entire region of study, but still concentrate in a few clearly defined groups. This led us to define some populations of stars, which we use as samples for our analysis. We named the two known clusters as DR15-C and DR15-N because their location center and north of our region of study. We identified and named as well three other groups: DR15-W, DR15-SW and DR15-SE. The purpose of this selection is to compare the age and age spread of these populations with the age of DR15-C. 

\subsubsection{YSO population ages from the K-band Luminosity Function }
\label{analysis:sfh:clusterages}

\par 
One of the goals of this study is to reconstruct the history of star formation in the region, by estimating the mean age and age spread of a set of the clusters and groups we identified from the \texttt{gather} maps. For this purpose, we constructed the K-band luminosity functions (KLF) of the groups and clusters, and compared them to the KLF of artificial pre-main sequence populations with different age ranges and age spreads. These artificial KLFs were constructed using the pre-main sequence model interpolation code of \citet{Muench:2000aa}. Another example of this method applied to a young cluster population can be found in \citet{Roman:2015aa}

\par We selected the samples for DR15-C,DR15-W, DR15-SW and DR15-SE for this analysis. The samples were defined as circular areas that covered the four overdensities. The radii of the circles was chosen as 0.03 deg for DR15-W,SW and SE, and as 0.0225 for DR15-C. The KLF was constructed by limiting the samples by extinction in the following manner: we restricted the sample to those sources that a) fall above an extinction vector corresponding $A_V$=20 mag that reaches the sensitivity limit in a $H$ vs. $J-K$ color-magnitude diagram, and b) fall between an unreddened and unreddened 3 Myr pre-main sequence isochrone, properly shifted to the estimated distance to DR15. The extinction-limited samples assure a minimum contamination from extragalactic sources (which mostly will fall in the area below the extinction vector and the sensitivity limit) and will minimize a bias due to the decrease in the number of detected sources as a function of extinction.

\par In their study, \citet{Muench:2000aa} showed that the observed shape of the KLF (mostly defined by the peak value) of a pre-main sequence population is particularly sensitive to three intrinsic parameters: the underlying Initial Mass Function (IMF), the mean age of the population, and to a lesser extent, the age spread or period of formation. Our analysis consists of constructing a grid of artificial KLF for each of our samples, and compare them with our observed function. We constructed this grid by assuming a fixed IMF and let the time parameters, mean age and age spread, to run free. 

\par For each population (DR15-C,DR15-W, DR15-SW and DR15-SE) we the used code of \citeauthor{Muench:2000aa} to simulate artificial KLFs for clusters with the same number of sources and the same distributions for extinction and disk fractions as a function of color ($H-K$). We drew the artificial population from a broken power law IMF using the parameters of the one obtained for the IC 348 cluster (2 Myr old) in the paper of \citet{Muench:2003aa}. To draw the artificial populations we used the  pre-main sequence stellar evolution model by \cite{DAntona:1997aa}, with $[D/H]=2\times 10^{-5}$ to draw the samples for the artificial clusters . We ran the models using a grid of ages in which we varied the mean age of the cluster between 0.5 and 10 Myr, in s.pdf of 0.5 Myr. For each case, we simulated five age spreads between 1.0 to 5.0 Myr. For each case we simulated 500 clusters.

The method could not be applied succesfully for the DR15-N sample, due to the large extinction variations in  that cluster, which does not allow us to obtain a sufficient detection rate in the K band at the highest column density regions. In the remaining clusters, we were able to isolate extinction limited samples satisfactorily. 

\begin{figure}
\includegraphics[width=3.0in]{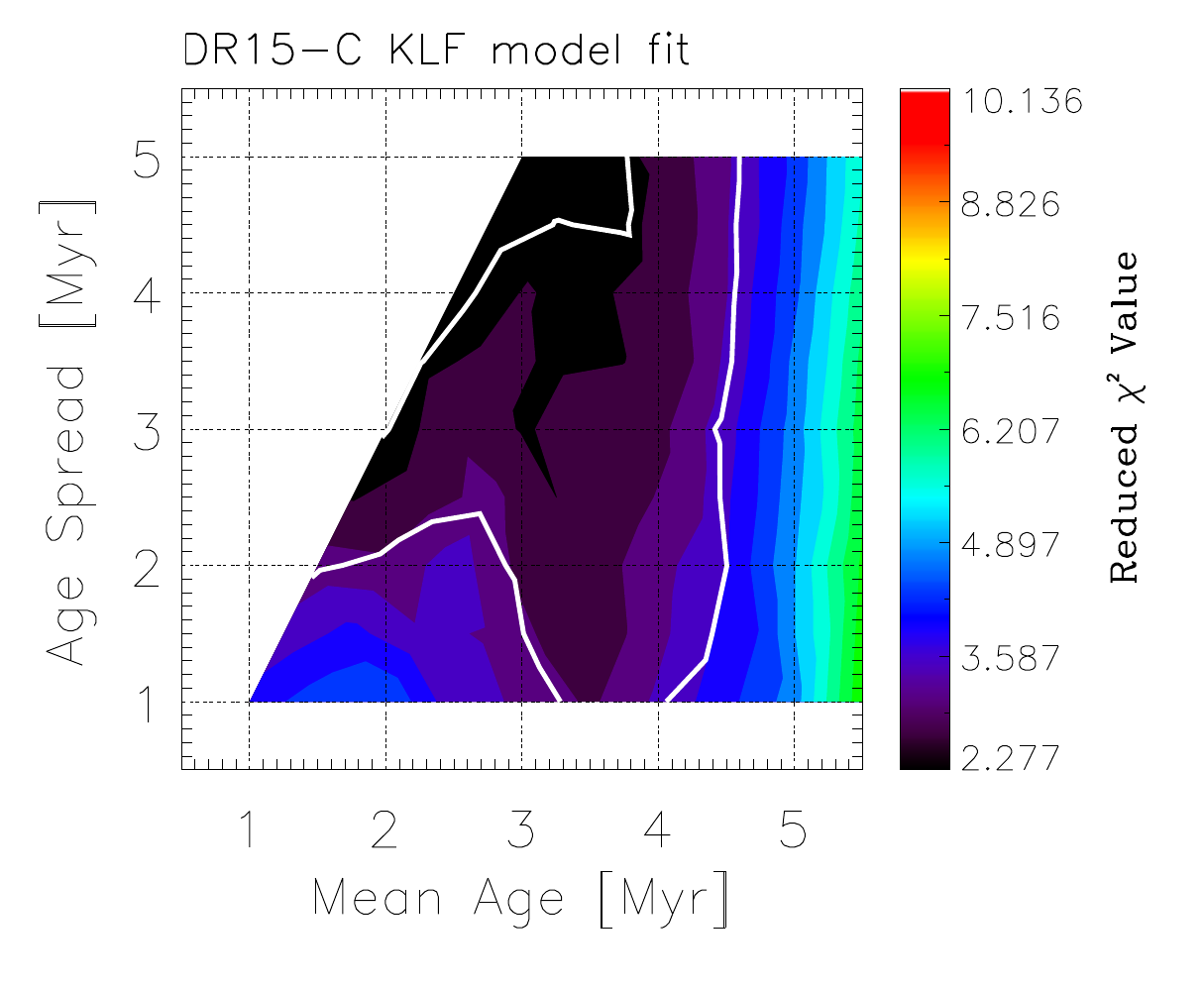}
\includegraphics[width=3.0in]{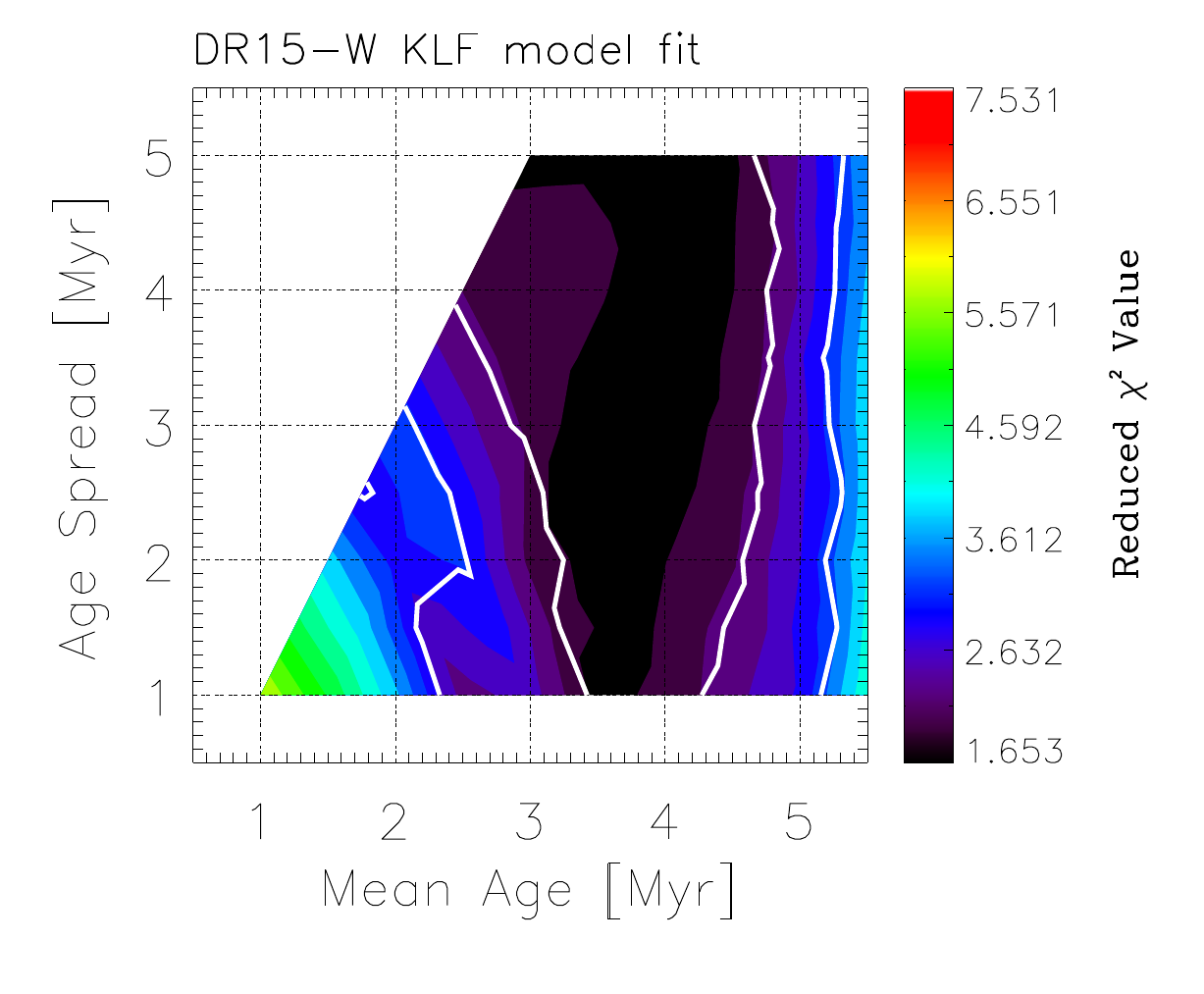}
\caption{The panels show contour maps of the normalized $\chi ^2$ values for the age vs. age spread estimation of clusters DR15-C and DR15-W using pre-main sequence models of the KLF. The most likely values fall within regions with purple and dark blue colors. The white contours indicate 68 and 95 percent confidence limits.The limits of the model grid are indicated by a thin, red line.}
\label{fig:nchi1}
\end{figure}

\begin{figure}
\includegraphics[width=3.0in]{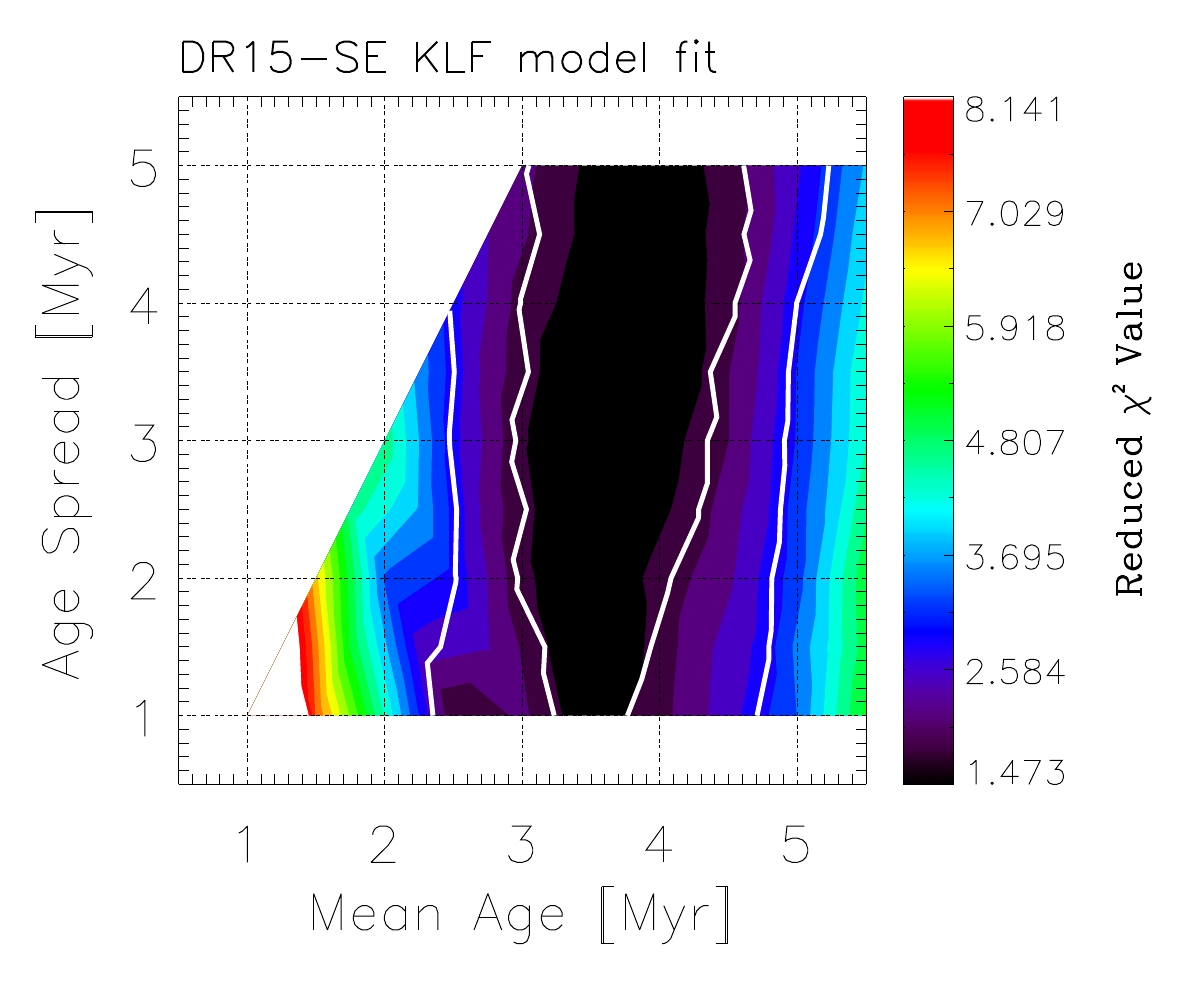}
\includegraphics[width=3.0in]{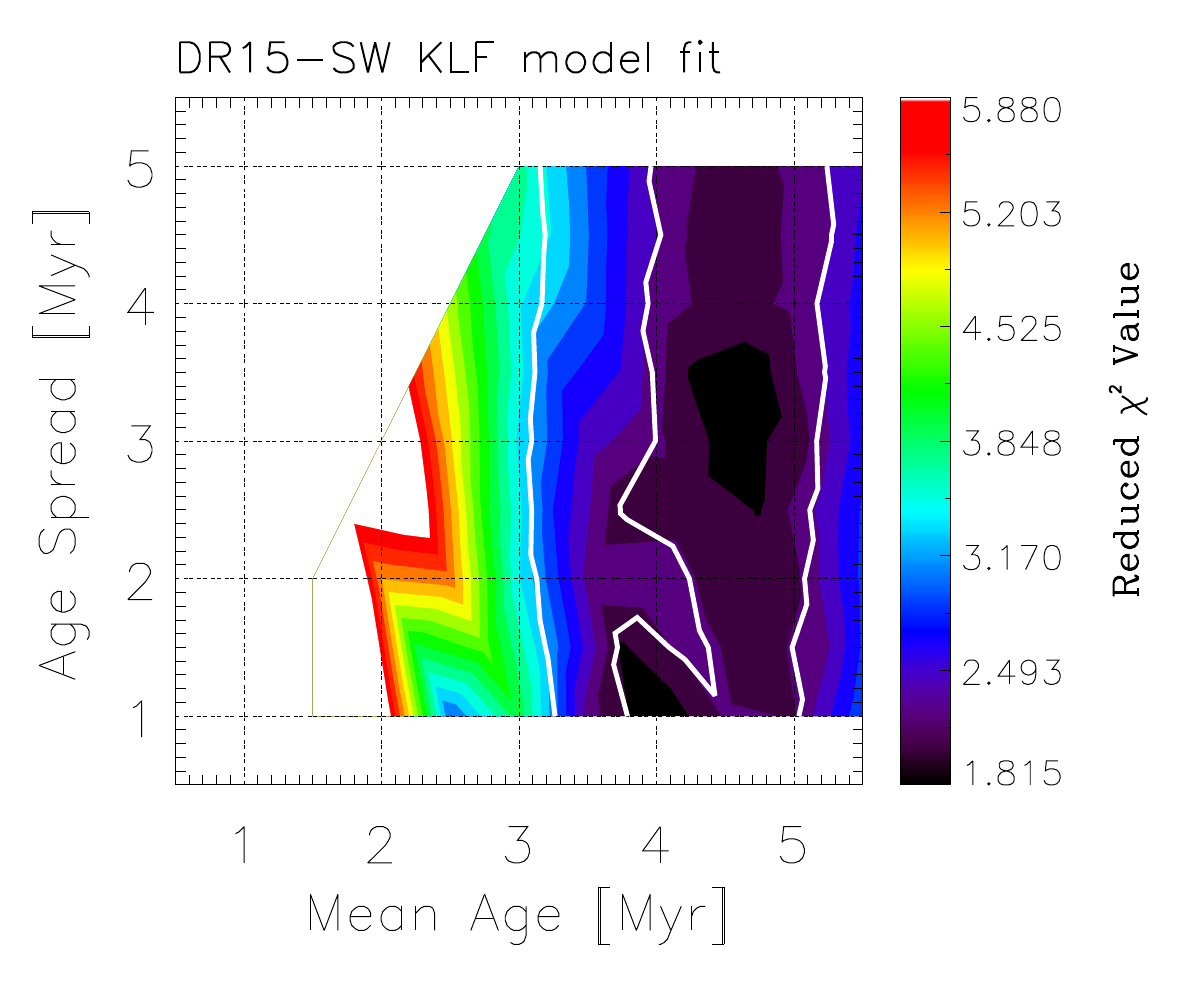}
\caption{Same as in Figure \ref{fig:nchi1}, but for clusters DR15-SW and DR15-SE}
\label{fig:nchi2}
\end{figure}

We determined which age/age spread set fits each of the observed KLFs by averaging the KLF over all simulations for each point in the grid and comparing to the observed KLF, obtaining for each case a reduced $\chi^2$ estimation. In Figures \ref{fig:nchi1} and \ref{fig:nchi2} we show, for each cluster a contour plot of the reduced $\chi^2$ in a mean age vs age spread plane, indicating the parameter region with the best adjustments.

%\clearpage

In Table \ref{tab:results} we list the results of the artificial KLF modeling analysis for the cluster samples in DR15. For each cluster we also include the number of Class I, Class II and Class III YSOs found in each population. The more embedded clusters DR15-C and DR15-W contain a population that according to our analysis, could have a mean age of 3.0 Myr. The populations flanking the main molecular pillar, DR15-SE and DR15-SW, appear to have mean ages as large as 3.5 and 4.5 Myr, respectively. The contour plots for these two samples, however, appear to have two local minima, suggesting age spreads as short as 3.0-3.5 Myr or as large as 4.5-5.0 Myr. The adjustment to an older age is in agreement with the less embedded status of these two samples. Notice that the confidence ranges in all cases indicate that our method cannot really constrain the age spreads satisfactorily (the 95 percent confidence range in DR15-C is the only one that suggest a constraint towards an age spread of 3.5 to 5 Myr). Still, given the embedded nature of these young star populations, we think it is little plausible that age spreads can be significantly larger than 5 Myr. 

\begin{center}
\begin{deluxetable}{lccccc}
\tablecolumns{6}
\tablewidth{0pt}
\tablecaption{Main Parameters and Age Estimates for Cluster Samples in the DR15 Region}
\tablehead{
\colhead{Cluster}  & 
\colhead{Mean Age\tablenotemark{a} } &
\colhead{Age Spread\tablenotemark{a}} & 
\colhead{No. CI\tablenotemark{b}} &   
\colhead{No. CII\tablenotemark{b}} & 
\colhead{No. CIII\tablenotemark{b}} \\
\colhead{} &
\multicolumn{2}{c}{[Myr]} &
\colhead{} &
\colhead{} &
\colhead{} \\
}
\startdata
DR15-C   & 3.0[2.5,3.5] & 4.5[3.5,5.0]  & 1  & 8  & 4  \\
DR15-W   & 3.5[2.5,4.5] & 4.5[1.0,5.0]  & 2  & 16 & 5  \\
DR15-SE  & 3.5[3.0,4.5] & 3.0[1.0,5.0]  & 0  & 5  & 0  \\
DR15-SW  & 4.5[4.0,5.0] & 3.0[1.0,5.0]  & 0  & 11 & 2  \\
DR15-N   & --           & --            & 15 & 21 & 6  \\
DR15-NE  & --           & --            & 0  & 11 & 3  \\
\enddata

\tablenotetext{a}{
The number pairs inside the brackets indicate the range of
model ages within the 2-sigma (68\%) confidence level. The number listed
before the bracket is the central value, which we list as the best estimate
for the age/age spread of the cluster}

\tablenotetext{b}{Estimated as number of sources of indicated
class falling within the circle used for the KLF sample analysis}

\label{tab:results}
\end{deluxetable}
\end{center}

%\clearpage

\subsection{The slow removal of the DR15-C cluster molecular envelope}
\label{analysis:sfh:radio_envelope}

\par At 3 Myr of age, we could expect that the DR15-C cluster have removed a significant fraction of its molecular envelope, as it occurs in clusters of a similar age, e.g. IC-348 \citep{Muench:2003aa}, IC 1795 \citep[e.g.][\ Rom\'an-Z\'u\~niga et al 2015, in rev.]{Oey:2005ly}. In fact our images show how other groups adjacent to DR15-C, like DR15-SE and DR15-SW, which our analysis suggests have similar ages, are associated with much less prominent molecular cloud features. Still, the molecular envelope of DR15-C appears as a well defined structure, both dense and compact, surrounding the cluster atop a dense molecular pillar.

\par Using the $^{13}$CO(1-0) map of \citet{Schneider:2011aa} we made two position-velocity (PV) cuts across the envelope of DR15-C. We used our extinction map and a zero moment (integrated intensity) integration of the $^{13}$CO(1-0) map as a guide. The first cut (L2) runs across the observable structure of the pillar in the region observed. The second cut (L4) runs almost perpendicular to L2. The L2 cut shows a component related to the neck of the pillar, almost 2 pc long with a radial velocity about 4 km/s away from the Cygnus OB2 system velocity (0 km/s). Near the center of the cut, coincidental with the envelope of DR15-C, the PV plot shows a hint of an elliptical shell structure, with a red component moving slightly above 4 km/s and a blue component near 0 km/s. The latter merges into the dark infrared cloud, which shows a very smooth gradient from 2 to 0 km/s. The L4 cut shows for the most part, gas with velocities near 0 km/s but also a much more clear elliptical shell structure at the envelope, which opens from 0 to 5 km/s. The elliptical shell structure corresponds to the expanding envelope of DR15-C.

\begin{figure}
\begin{centering}
\includegraphics[width=3.0in]{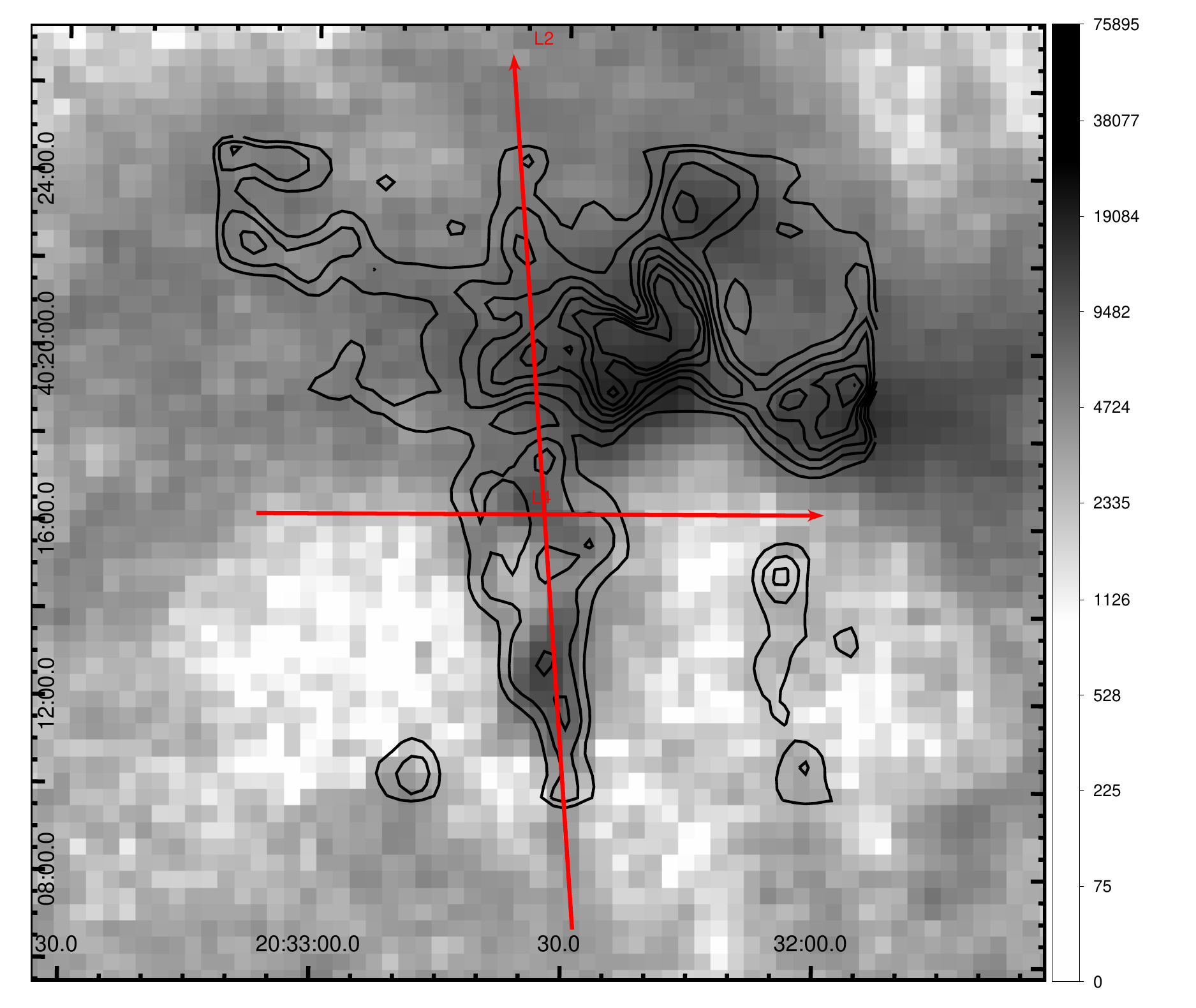}\\
\includegraphics[width=3.0in]{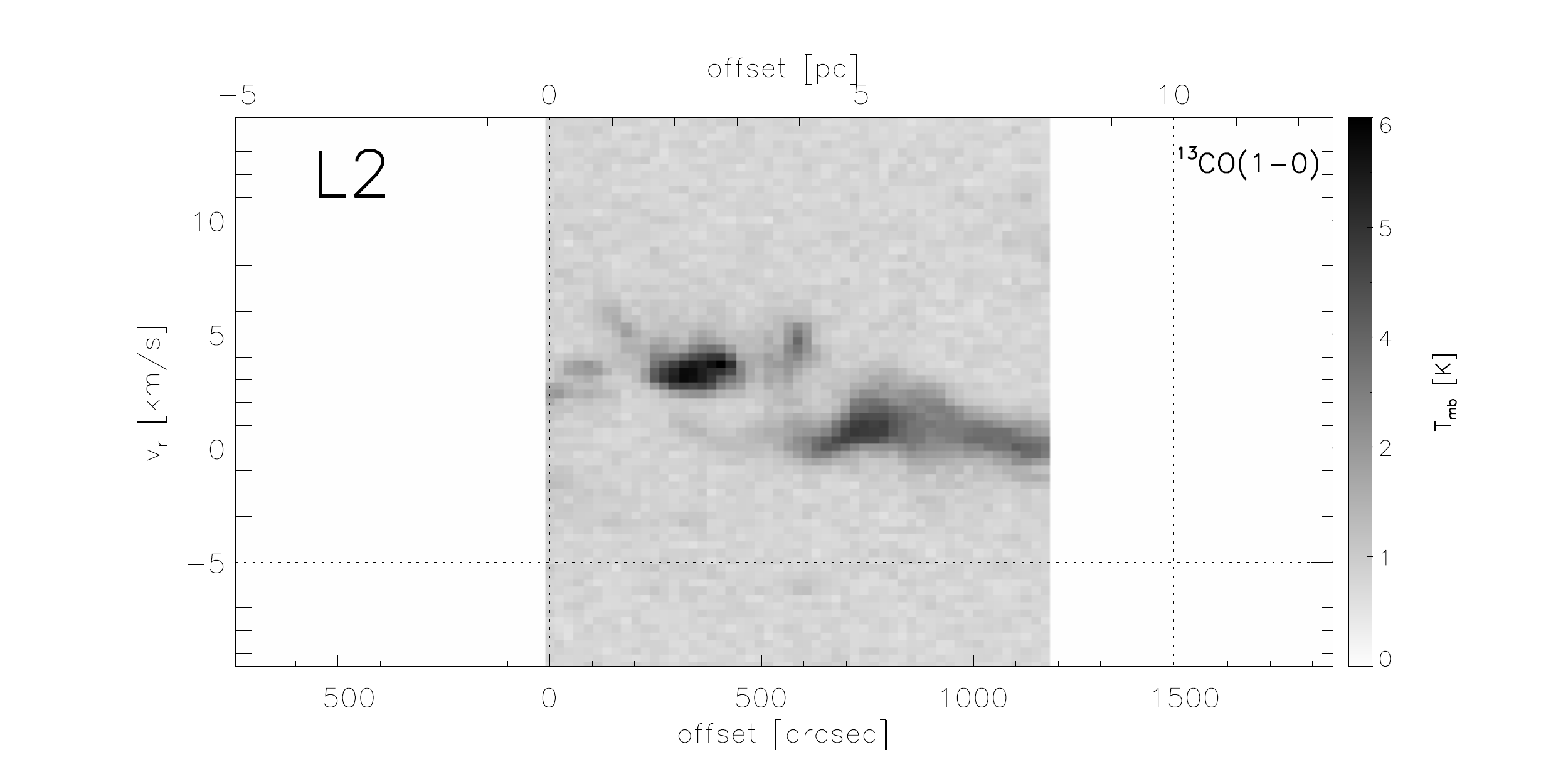}\\
\includegraphics[width=3.0in]{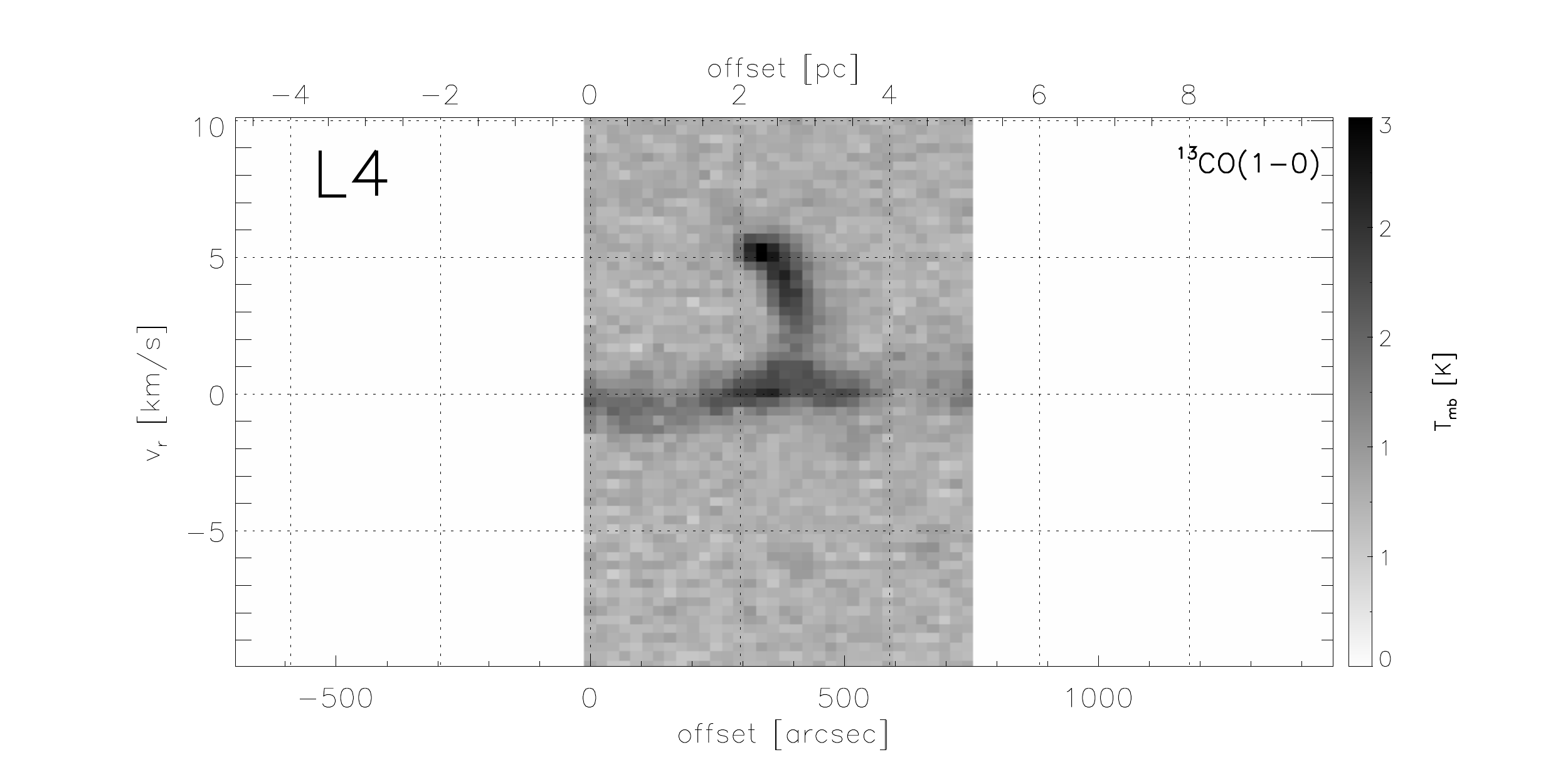}
\caption{Top: Zero moment (integrated intensity) $^{13}$CO(1-0) map of the DR15 region, constructed from the survey of \citet{Schneider:2006aa}, the grayscale is indicated in units of km/s. Contours indicate visual extinction, similar to Figure \ref{fig:ysodist}. The two red arrow lines labeled as L2 and L4 indicate cuts along which position-velocity (P-V) plots were obtained (see text). Center, Bottom: P-V plots along L2 and L4 cuts, as indicated in top panel; position along the length of each cut is indicated both in arcseconds and in parsecs, asumming a distance of 1.4 kpc.}
\label{fig:pvcuts}
\end{centering}
\end{figure}
%\clearpage

\par We defined the area of the shell as a rectangle of 5.25$\arcmin \times$3.75$\arcmin$ around the defined center of DR15C. Following the prescription by \citet{Estalella:1999aa}, we estimated the column density, N(H$_2$) and the total mass of expanding gas within $3<v_r<6$ km/s, as $M_{out}=$103.5 M$_\odot$. Then we used the method described by \citep{Qiu:2009aa} to estimate the dynamical time of the component, $t_{dyn}=2.9\times 10^5$ yr. This implies a total mass loss rate of $\dot{M}_{out}=M_{out}/t_{dyn}=360 \mathrm{\ M}_\odot \mathrm{\ Myr}^{-1}$. At this rate, it would take about 3 Myr to remove the total mass of gas in the shell, which we estimate to be 1023 M$_\odot$, using our extinction map, and a distance to the pillar of 1.4 kpc.

\par This result appears to be consistent with estimates of the mass loss rates in other molecular pillars. For instance, using integral field unit spectroscopy, \citet{Westmoquette:2013aa} estimated a mass loss rate of 300 $\mathrm{\ M}_\odot \mathrm{\ Myr}^{-1}$ for the pillars of NGC 3603, which contain a total of about 700 M$_\odot$ of gas. For the ``Pillars of creation" in M16, an estimate by \citet{Mcleod:2015aa} is of 70 $\mathrm{\ M}_\odot \mathrm{\ Myr}^{-1}$. In both cases, the removal timescales are of the order of 2-3 Myr. However, we need to point out that those estimates are based on models of photoevaporation, while our estimate comes from an estimation of a gas outflow rate from a zero moment map of molecular gas emission. The shell mass could be overestimated, due mainly to a large column density in the line of sight towards Cygnus-X, in which case we would be more consistent with the optical spectroscopy studies. However, what we consider important to note is that the removal of the envelope of DR15-C is relatively slow compared to other embedded cluster regions, where gas dispersal timescales are similar or shorter than the T Tauri timescale, i.e. less than 2 Myr  .

\section{Discussion }
\label{discussion}

\par The main goal of this paper is to reconstruct the history of star formation in the Cygnus-X DR15 region. For this purpose, we made a) an analysis of the spatial distribution of YSO candidate sources classified by evolutionary classes, and b) a comparison of the observed KLF of several young star population samples in our region of study with those of artificial cluster samples drawn from pre-main sequence stellar evolution models.

\par The classification of YSO candidates and their spatial distribution reveals that several populations of young stars are present in the field of study. At the north/northeast part of our field we identified a very young cluster, DR15-N, hosting 15 Class 0/I sources. The cluster is forming within an infrared dark cloud that runs in the east-west direction at an estimated distance of 1400 pc \citep{Rizzo:2014aa,Palau:2014aa}. The cluster also contains a relatively large number (21) of Class II sources and thus it may host the youngest population in the region of study. For this cluster we were not able to construct an unbiased extinction-limited sample to construct a KLF. Instead, we constructed SEDs for all the Class I sources we identified within it, and we compared them to YSO models. We found that in most cases, the models with the best fits correspond to intermediate to high mass YSOs with ages of $\sim 1$ Myr. The mass of the IRDC, estimated from our NICEST extinction map is of $\sim 2400\mathrm{\ M}_\odot$, with an equivalent radii of 5.2 pc. These parameters are in good agreement with massive star forming IRDCs as defined by the analysis of \citet{Kauffmann:2010aa}.

\par The revised distance to Cygnus-X from the study of \citet{Rygl:2012aa} is 1.4 kpc, in agreement with the estimations of \citet{Rizzo:2014aa} and \citet{Palau:2014aa}. From our position-velocity cuts, we note that the emission along the structure of the dark infrared cloud has a null relative velocity compared to the Cygnus-X systemic velocity, same as the DR12-15 structure. It would be difficult to determine if the dark infrared cloud is located at the same distance than the DR15 pillar based only on a radial velocity difference argument. Also, it is important to notice that \citeauthor{Rygl:2012aa} could not confirm that Cygnus-X South structures are at the same distance as those in Cygnus-X North where most of their measurements were made. We also lack enough evidence to claim interaction between the DR15 pillar and the IRDC. Using purely morphological arguments we like to comment that the formation of molecular pillars like DR15 is thought to be the result of the interaction between the molecular cloud and the photoionization region formed by the ionizing radiation of the central cluster (Cygnus OB2 in our case), while the IRDC appears to be a highly dense and coherent structure, apparently less affected by the HII regions. Also, our images do not show any obvious pillar structures coincident with the IRDC. These arguments could work in favor of a scenario in which the IRDC could be located at a slightly different distance than the DR15 pillar.

\par According to our observations and our KLF simulation results, DR15-C is a young cluster with an age of about $~3$ Myrs, but it has a low number of young sources, with only one identified Class I, 8 Class II and 4 Class III  sources, all of this despite the presence of the thick, nebulous envelope evident in extinction and $^{13}$CO(1-0) integrated intensity maps. We know the DR15 pillar hosts an embedded cluster, as evidenced by well known far-infrared sources associated with the photodissociation region \citep{Odenwald:1990aa} and also from our KLF analysis, which indicates that 200 sources could be present within DR15-C, after background and foreground contamination corrections. We do not think the contamination by sources from the DR15-N cluster is too high, as most the density peak of that cluster is located to the northeast and it is more embedded. However, it cannot be fully discarded that some sources in DR15-C are actually sources from DR15-N (and viceversa). Also, there seems to be another YSO group in the filament, south of DR15-C, which reinforces the idea of star formation being active in the pillar. 

\par Our evidence points to a scenario in which a) the star formation episode in DR15-C has probably reached its end or it is near its end and b) the parental gas has dissipated relatively slow, or at least slower that the dispersion time of circumstellar disks of its member stars. Our analysis of the $^{13}$CO(1-0) zero moment map indicates it would take up to 4 Myr to remove the cluster gaseous envelope. However, our observations in other cluster forming regions like the Rosette Molecular Cloud and W3 \citep[e.g.][]{Roman-Zuniga:2008aa,Ybarra:2013aa,Roman:2015aa} suggest that clusters hosting intermediate to massive stars may remove their gas envelopes in periods shorter than the T Tauri timescale. Therefore, our analysis of DR15-C suggests that clusters forming in this kind of pillar structures, could dissipate their envelopes at slower rates. 

As shown in recent studies like those of \citet{Westmoquette:2013aa} and \citet{Mcleod:2015aa} the mass loss rates in pillars due to photoevaporation are of the order of $\sim 10^2\mathrm{\ M}_\odot \mathrm{\ Myr}^{-1}$, not too different from our estimations in the expanding gas shell. Clearly, we are comparing very different methods, that refer to very different processes (shell expansion vs. shell photo-erosion). It is important to notice that the DR15 pillar may not be the same kind of dusty pillar as those in M16, because extinction is too high to allow delineating the structure in DR15 using optical images. The molecular fragment located west of DR15 shows as a shadow with a pillar morphology in optical images, which make us think that DR15 could be a pillar too. Even so, the fact that the mass loss estimates for DR15 coincide in the order of magnitude of the effect with respect to M16 is interesting, and motivates further investigation. A detailed study of the removal of the DR15-C envelope is out of the scope of this paper, but it is the main topic in a close following study (Rom\'an-Z\'u\~niga et al., in prep).

\par DR15-W, DR15-SE and DR15-SW appear to be slightly older populations ($3.5-4.75$ Myrs) with much less gas and dust observed.  However, we find evidence of remains of a structure that could have been similar to the pillar associated with DR15-C. It may be possible that the DR15-SE and DR15-SW groups belong to clusters formed before DR15-C, but their ages may not be much older, as evident from the presence of Class II and Class III sources. The estimated ages of the DR15-SW and SE samples suggest that the cluster evolution period, from formation to gas dispersal in the region could be around 5 Myr. 

\section{Summary}
\label{summary}

\par  The Cygnus-X DR15 region presents a prominent gaseous pillar as well as an IRDC, both hosting clusters of young stars. In this investigation, we made a multi-wavelength study of the young stellar population in the region. For this purpose, we processed and analyzed deep, high quality near-IR images of the region, as well as X-ray images from the Chandra Observatory. Using these datasets we obtained photometry catalogs for all point sources we were able to detect. We combined these catalogs with the 3.6 to 24 $\mu$m photometry catalog of the Cygnus-X Spitzer Legacy Survey, resulting in a master catalog containing almost 47 thousand individual sources. From our master catalog, we identified 226 young stellar sources, which we classified according to their evolutionary class related to the prominence of their circumstellar disks. We found that the young sources distribute into 26 Class 0/I, 156 Class II and 45 Class III sources.

\par From our near-IR we constructed an extinction map, which we used to study the spatial distribution of the young sources in the molecular cloud structure present in the region. We found that the youngest population of this region is currently forming at the IRDC, at the regions of highest column density. Combining this with maps of YSO surface density, we were able to identify several groups, possibly associated to distinct stellar clusters. We obtained extinction-limited samples of these groups in order to construct their K band luminosity functions (KLF). We compared the observed KLFs with those of artificial young cluster populations sampled from interpolation of pre-main sequence models. This allowed us to make first order estimations of the mean ages and age spreads of the cluster population samples.

\par We constructed SEDs for all Class I sources identified within the IRDC region, which allowed us to estimate their mass and disk accretion rates. These estimations are consistent with the formation of an intermediate mass star cluster, indicating that structures of this kind in Cygnus-X are possibly birth places of massive clusters. 

\par Using the FCRAO $^{13}$CO(1-0) from the study of \citet{Schneider:2011aa} we estimated the radial velocity distribution along the IRDC and the DR15 pillar, and we found that the nebulous envelope of the DR15-C cluster at the tip of the pillar is consistent with an expanding shell morphology. We estimated the mass loss rate of gas in this expanding shell and we deduced that the dissipation process of the DR15-C cluster gas envelope is relatively slow, compared to what we found in other studies of cluster forming regions, where the gas dispersal process is shorter than the T Tauri timescale. The mass loss rate we estimated is in the same order of magnitude as mass loss rates by photodissociation found in studies of other gas pillars using optical spectroscopy. This suggests that clusters forming in gas pillars like DR15-C could have a different evolution process than clusters forming at dense clumps in other giant molecular clouds. 

\par The presence of other populations containing Class II and Class III sources at the regions flanking the DR15 pillar, and projected near less dense but still noticeable gaseous structures, with estimated ages of 4 to 5 Myr is suggestive of a process of cluster forming processes that take about that long to form an dissipate in the molecular cloud complex that surrounds the Cygnus-X HII region.

\acknowledgements

\par \par \textbf{Acknowledgements:} We thank the referee, Nicola Schneider for providing a comprehensive and constructive review that greatly improved the content of our manuscript. We thank Nicola Schneider for kindly providing us with a copy of the FCRAO  $^{13}$CO(1-0) map for our study. CRZ acknowledges support from CONACYT project CB2010-152160, Mexico and programs UNAM-DGAPA-PAPIIT IN103014 and IN116315. EAL acknowledges support from the National Science Foundation through NSF Grant AST-1109679 to the University of Florida.

\par This study is based on observations collected at the Centro Astron\'omico Hispano Alem\'an (CAHA) at Calar Alto, operated jointly by the Max-Planck Institut f\"ur Astronomie and the Instituto de Astrof\'isica de Andaluc\'ia (CSIC). We acknowledge the staff at Calar Alto for top of the line queued observations at the 3.5m with OMEGA 2000. We acknowledge use of data products from the 2MASS, which is a joint project of the University of Massachusetts and the Infrared Processing and Analysis Centre/California Institute of Technology (funded by the USA National Aeronautics and Space Administration and National Science Foundation). This work is partly based on observations made with the Spitzer Space Telescope, which is operated by the Jet Propulsion Laboratory, California Institute of Technology under a contract with NASA. The scientific results reported in this article are based to a significant degree on data obtained from the Chandra Data Archive; particularly, we made use of data obtained from the Chandra Source Catalog, provided by the Chandra X-ray Center (CXC) as part of the Chandra Data Archive. We made use of the \texttt{pvextractor} tool by Adam Ginsburg, that is part of the Radio Astro Tools repository (http://github.com/radio-astro-tools).

{\it Facilities:} \facility{CAO:3.5m (OMEGA2000)}, \facility{Spitzer (IRAC,MIPS)}, \facility{CXO (ACIS)}.

\appendix

\section{Young Sources Identified in the DR15 region \label{app:ysos}}

\par In Tables \ref{tab:ClassI}, \ref{tab:ClassII} and \ref{tab:ClassIII} we list YSO sources identified as Class I, Class II and Class III in our region of study. The tables contain JHK photometry from Calar Alto (or 2MASS when pertinent, see section \ref{obs:nir}), 3.6 to 24 $\mu$m photometry from the Spitzer CXLS, and when possible, median energy and total energy flux values for those sources detected with Chandra ACIS (that is the case for all Class III sources).
Identifications and positions from our Calar Alto survey are listed for most cases. The remaining sources are listed with 2MASS or CXLS depending if they were detected in those surveys.

\clearpage
\begin{landscape}

\begin{deluxetable}{lllllllllllllllllllll}
\tabletypesize{\tiny}
\setlength{\tabcolsep}{0.015in}
\tablecolumns{21}
\tablewidth{0pt}
\tablecaption{Class I sources in the DR15 regionn}
\tablehead{
\colhead{Source}  & 
\colhead{RA} &
\colhead{DEC} &
\colhead{J} &
\colhead{$\sigma_J$} &
\colhead{H} &
\colhead{$\sigma_H$} &
\colhead{K} &
\colhead{$\sigma_K$} &
\colhead{[3.6]} &
\colhead{$\sigma_{[3.6]}$} &
\colhead{[4.5]} &
\colhead{$\sigma_{[4.5]}$} &
\colhead{[5.8]} &
\colhead{$\sigma_{[5.8]}$} &
\colhead{[8.0]} &
\colhead{$\sigma_{[8.0]}$} &
\colhead{[24\ $\mu$m]} &
\colhead{$\sigma_{24\mu m}$} &
\colhead{Median Energy} &
\colhead{Energy Flux} \\
\colhead{} &
\multicolumn{2}{c}{[J2000]} &
\multicolumn{16}{c}{[mag]} &
\colhead{[keV]} &
\colhead{[erg/cm$^2$/s]} \\
}
\startdata

  CAHA\_20322208\_402017 & 20:32:22.08 & +40:20:17.2 & 18.827 & 9.999 & 18.101 & 0.02 & 12.026 & 0.02 & 7.411 & 0.015 & 6.021 & 0.015 & 5.066 & 0.015 & 4.136 & 0.015 & 0.395 & 0.018 & --- & ---\\
  CAHA\_20322856\_401941 & 20:32:28.56 & +40:19:41.6 & 14.827 & 0.003 & 11.397 & 0.019 & 8.943 & 0.016 & 7.372 & 0.015 & 6.304 & 0.015 & 5.349 & 0.015 & 4.427 & 0.015 & 0.981 & 0.017 & --- & ---\\
  CAHA\_20321777\_401408 & 20:32:17.77 & +40:14:08.2 & 15.139 & 0.015 & 12.744 & 0.014 & 10.647 & 0.014 & 8.162 & 0.015 & 7.323 & 0.015 & 6.455 & 0.015 & 4.864 & 0.015 & 1.412 & 0.016 & --- & ---\\
  SSTCYGX\_J203222.99\_402021.4 & 20:32:22.99 & +40:20:21.4 & --- & ---  & --- & ---  & --- & ---  & 13.706 & 0.17 & 11.445 & 0.106 & 9.627 & 0.041 & 7.858 & 0.033 & 1.662 & 0.019 & --- & ---\\
  CAHA\_20322060\_401950 & 20:32:20.60 & +40:19:50.1 & --- & ---  & --- & ---  & 16.815 & 0.02 & 10.233 & 0.021 & 8.048 & 0.016 & 6.613 & 0.015 & 5.777 & 0.017 & 2.145 & 0.038 & --- & ---\\
  CAHA\_20323152\_401352 & 20:32:31.52 & +40:13:53.0 & 20.457 & 0.042 & 14.909 & 0.008 & 14.374 & 0.089 & 8.519 & 0.016 & 7.417 & 0.015 & 6.565 & 0.017 & 5.96 & 0.03 & 2.229 & 0.111 & --- & ---\\
  CAHA\_20322113\_402025 & 20:32:21.13 & +40:20:25.6 & 17.814 & 9.999 & 16.994 & 0.013 & 13.068 & 0.009 & 9.281 & 0.016 & 8.148 & 0.016 & 7.201 & 0.016 & 6.35 & 0.017 & 2.316 & 0.021 & 2.694 & 5.880E-15\\
  CAHA\_20322194\_401937 & 20:32:21.94 & +40:19:37.9 & --- & ---  & 20.053 & 0.075 & 15.564 & 0.017 & 11.613 & 0.021 & 10.495 & 0.02 & 7.246 & 0.016 & 5.487 & 0.021 & 2.668 & 0.056 & --- & --- \\
  CAHA\_20322126\_401601 & 20:32:21.26 & +40:16:01.8 & 18.458 & 0.023 & 14.213 & 0.007 & 11.565 & 0.018 & 8.827 & 0.015 & 7.851 & 0.015 & 7.005 & 0.015 & 6.26 & 0.017 & 3.036 & 0.18 & --- & --- \\
  CAHA\_20320254\_401838 & 20:32:02.54 & +40:18:39.0 & --- & ---  & --- & ---  & 17.492 & 0.073 & 11.914 & 0.017 & 10.031 & 0.016 & 8.879 & 0.017 & 8.241 & 0.033 & 3.077 & 0.046 & --- & --- \\
  CAHA\_20322282\_401940 & 20:32:22.82 & +40:19:40.9 & --- & ---  & --- & ---  & 16.245 & 0.038 & 12.171 & 0.03 & 10.362 & 0.019 & 9.161 & 0.021 & 8.26 & 0.059 & 3.583 & 0.118 & --- & --- \\
  CAHA\_20315797\_401835 & 20:31:57.97 & +40:18:35.8 & 21.759 & 0.22 & --- & ---  & --- & ---  & 13.975 & 0.028 & 12.189 & 0.027 & 11.35 & 0.036 & 10.973 & 0.112 & 3.677 & 0.035 & --- & --- \\
  SSTCYGX\_J203153.84\_401833.9 & 20:31:53.84 & +40:18:33.9 & --- & ---  & --- & ---  & --- & ---  & 13.039 & 0.018 & 11.039 & 0.016 & 9.802 & 0.019 & 8.87 & 0.028 & 5.383 & 0.092 & --- & --- \\
  CAHA\_20322222\_401955 & 20:32:22.22 & +40:19:56.0 & 18.432 & 9.999 & 16.975 & 0.008 & 13.395 & 0.005 & 10.72 & 0.016 & 9.534 & 0.016 & 8.519 & 0.018 & 7.538 & 0.025 & --- & ---  & --- & --- \\
  CAHA\_20322014\_401953 & 20:32:20.14 & +40:19:53.6 & 17.771 & 9.999 & 17.345 & 9.999 & 14.847 & 0.006 & 9.945 & 0.016 & 8.612 & 0.016 & 7.726 & 0.017 & 7.577 & 0.026 & --- & ---  & --- & --- \\
  CAHA\_20322784\_401942 & 20:32:27.84 & +40:19:42.4 & 19.796 & 0.037 & 15.484 & 0.007 & 12.902 & 0.005 & 10.559 & 0.017 & 9.733 & 0.018 & 8.976 & 0.021 & 8.266 & 0.062 & --- & ---  & --- & --- \\
  CAHA\_20322033\_402001 & 20:32:20.33 & +40:20:01.5 & 16.173 & 0.003 & 15.529 & 0.007 & 15.118 & 0.012 & 13.54 & 0.118 & 11.84 & 0.066 & 10.647 & 0.059 & 9.852 & 0.075 & --- & ---  & --- & --- \\
  CAHA\_20322899\_401821 & 20:32:28.99 & +40:18:21.2 & --- & ---  & --- & ---  & 16.45 & 0.03 & 14.072 & 0.062 & 13.326 & 0.036 & 12.328 & 0.117 & 11.479 & 0.24 & --- & ---  & --- & --- \\
  CAHA\_20322111\_402001 & 20:32:21.11 & +40:20:01.1 & --- & ---  & --- & ---  & 16.686 & 0.047 & 11.988 & 0.033 & 10.318 & 0.023 & 9.153 & 0.019 & 8.38 & 0.027 & --- & ---  & --- & --- \\
  CAHA\_20322781\_402032 & 20:32:27.81 & +40:20:32.7 & --- & ---  & --- & ---  & 18.385 & 0.1 & 14.168 & 0.022 & 12.909 & 0.018 & 12.075 & 0.042 & 11.14 & 0.069 & --- & ---  & --- & --- \\
  SSTCYGX\_J203226.39\_401847.4 & 20:32:26.39 & +40:18:47.4 & --- & ---  & --- & ---  & --- & ---  & 14.761 & 0.051 & 13.612 & 0.028 & 12.252 & 0.112 & 11.323 & 0.185 & --- & ---  & --- & --- \\
  CAHA\_20320404\_401856 & 20:32:04.04 & +40:18:56.6 & 18.559 & 0.009 & 15.162 & 0.01 & 13.261 & 0.01 & 11.352 & 0.017 & 10.546 & 0.016 & 9.895 & 0.03 & 9.09 & 0.064 & 5.954 & 0.42 & 4.636 & 1.717E-14\\
  CAHA\_20323480\_401629 & 20:32:34.80 & +40:16:29.2 & 17.797 & 0.023 & 14.591 & 0.016 & 12.513 & 0.009 & 10.185 & 0.05 & 9.455 & 0.048 & 8.048 & 0.145 & --- & ---  & --- & ---  & 3.906 & 1.154E-14\\
  CAHA\_20322210\_401800 & 20:32:22.10 & +40:18:00.3 & 13.681 & 0.003 & 12.608 & 0.006 & 12.447 & 0.023 & 11.337 & 0.021 & 11.004 & 0.022 & 10.653 & 0.153 & 9.602 & 0.323 & --- & ---  & 1.438 & 6.480E-15\\
  CAHA\_20322358\_401729 & 20:32:23.58 & +40:17:29.9 & 14.661 & 0.003 & 13.556 & 0.005 & 13.146 & 0.005 & 12.243 & 0.08 & 12.081 & 0.112 & 9.757 & 0.146 & --- & ---  & --- & ---  & 1.204 & 4.064E-15\\
  CAHA\_20325161\_401945 & 20:32:51.61 & +40:19:45.6 & 16.385 & 0.005 & 14.341 & 0.007 & 13.271 & 0.005 & 12.165 & 0.017 & 11.689 & 0.017 & 11.375 & 0.024 & 11.162 & 0.054 & --- & ---  & 3.351 & 1.147E-14\\
  
\enddata
\label{tab:ClassI}
\end{deluxetable}

\clearpage
\begin{deluxetable}{lllllllllllllllllllll}
\tabletypesize{\tiny}
\setlength{\tabcolsep}{0.015in}
\tablecolumns{21}
\tablewidth{0pt}
\tablecaption{Class II sources in the DR15 regionn}
\tablehead{
\colhead{Source}  & 
\colhead{RA} &
\colhead{DEC} &
\colhead{J} &
\colhead{$\sigma_J$} &
\colhead{H} &
\colhead{$\sigma_H$} &
\colhead{K} &
\colhead{$\sigma_K$} &
\colhead{[3.6]} &
\colhead{$\sigma_{[3.6]}$} &
\colhead{[4.5]} &
\colhead{$\sigma_{[4.5]}$} &
\colhead{[5.8]} &
\colhead{$\sigma_{[5.8]}$} &
\colhead{[8.0]} &
\colhead{$\sigma_{[8.0]}$} &
\colhead{[24\ $\mu$m]} &
\colhead{$\sigma_{24\mu m}$} &
\colhead{Median Energy} &
\colhead{Energy Flux} \\
\colhead{} &
\multicolumn{2}{c}{[J2000]} &
\multicolumn{16}{c}{[mag]} &
\colhead{[keV]} &
\colhead{[erg/cm$^2$/s]} \\
}
\startdata
  CAHA\_20324212\_401726 & 20:32:42.12 & +40:17:27.0 & 12.732 & 0.005 & 11.778 & 0.03 & 11.309 & 0.023 & 11.001 & 0.018 & 10.834 & 0.018 & 10.6 & 0.103 & 9.503 & 0.22 & 2.175 & 0.175 & --- & --- \\
  TWOM\_20322301\_4019226 & 20:32:23.02 & +40:19:22.7 & 12.204 & 0.025 & 10.916 & 0.021 & 10.071 & 0.014 & 8.789 & 0.015 & 8.154 & 0.015 & 7.561 & 0.015 & 6.622 & 0.016 & 2.94 & 0.037 & --- & --- \\
  CAHA\_20323244\_402105 & 20:32:32.44 & +40:21:05.1 & 13.473 & 0.003 & 11.578 & 0.018 & 10.661 & 0.017 & 10.131 & 0.015 & 9.936 & 0.015 & 9.751 & 0.016 & 9.249 & 0.024 & 3.358 & 0.043 & --- & --- \\
  CAHA\_20321111\_401916 & 20:32:11.11 & +40:19:16.7 & 19.311 & 0.033 & 15.549 & 0.016 & 13.14 & 0.007 & 11.057 & 0.03 & 9.915 & 0.023 & 9.281 & 0.096 & 8.539 & 0.248 & 3.658 & 0.235 & --- & --- \\
  CAHA\_20321736\_401949 & 20:32:17.36 & +40:19:49.5 & 19.194 & 0.02 & 15.036 & 0.006 & 12.555 & 0.004 & 10.287 & 0.017 & 9.536 & 0.016 & 8.877 & 0.063 & 8.238 & 0.205 & 3.748 & 0.159 & --- & --- \\
  CAHA\_20320319\_402215 & 20:32:03.19 & +40:22:15.9 & 14.418 & 0.004 & 12.749 & 0.008 & 12.095 & 0.02 & 9.921 & 0.016 & 9.256 & 0.015 & 8.651 & 0.017 & 7.743 & 0.038 & 4.189 & 0.086 & --- & --- \\
  CAHA\_20324852\_402104 & 20:32:48.52 & +40:21:04.8 & 13.849 & 0.003 & 12.805 & 0.005 & 12.1 & 0.021 & 10.642 & 0.015 & 9.731 & 0.015 & 9.054 & 0.016 & 8.221 & 0.021 & 4.79 & 0.046 & --- & --- \\
  CAHA\_20325574\_402220 & 20:32:55.74 & +40:22:20.6 & --- & ---  & 19.872 & 0.06 & 16.277 & 0.041 & 11.28 & 0.016 & 9.469 & 0.015 & 8.207 & 0.015 & 7.367 & 0.015 & 4.826 & 0.036 & --- & --- \\
  CAHA\_20315532\_402216 & 20:31:55.32 & +40:22:16.8 & 14.671 & 0.003 & 13.231 & 0.008 & 12.382 & 0.006 & 11.501 & 0.016 & 11.015 & 0.016 & 10.594 & 0.019 & 9.544 & 0.026 & 4.841 & 0.196 & --- & --- \\
  CAHA\_20321148\_401807 & 20:32:11.71 & +40:18:05.1 & 13.322 & 0.039 & 12.351 & 0.042 & 11.517 & 9.999 & 10.704 & 0.019 & 10.358 & 0.018 & 10.261 & 0.051 & 9.379 & 0.117 & 4.949 & 0.169 & --- & --- \\
  CAHA\_20320833\_401604 & 20:32:08.33 & +40:16:04.7 & 13.914 & 0.004 & 12.784 & 0.005 & 12.356 & 0.023 & 11.52 & 0.019 & 11.226 & 0.017 & 10.55 & 0.061 & 9.419 & 0.111 & 5.031 & 0.162 & 1.774 & 1.365E-14\\
  CAHA\_20321289\_401257 & 20:32:12.89 & +40:12:58.0 & 14.167 & 0.011 & 12.895 & 0.021 & 12.215 & 0.02 & 11.591 & 0.016 & 11.367 & 0.016 & 11.211 & 0.04 & 10.694 & 0.089 & 5.4 & 0.076 & 2.069 & 9.625E-15\\
  CAHA\_20324359\_402121 & 20:32:43.59 & +40:21:21.3 & 14.853 & 0.003 & 12.584 & 0.007 & 11.295 & 0.017 & 9.688 & 0.015 & 9.134 & 0.015 & 8.61 & 0.015 & 7.939 & 0.016 & 5.42 & 0.07 & --- & --- \\
  TWOM\_20315270\_4019059 & 20:31:52.71 & +40:19:06.0 & 17.14 & 0.217 & 14.15 & 0.042 & 12.799 & 0.029 & 12.079 & 0.019 & 11.8 & 0.026 & 11.499 & 0.045 & 10.641 & 0.084 & 5.502 & 0.245 & 2.373 & 9.074E-15\\
  CAHA\_20315661\_401614 & 20:31:56.61 & +40:16:14.0 & 12.979 & 0.005 & 12.093 & 0.019 & 11.357 & 0.016 & 10.733 & 0.016 & 10.412 & 0.016 & 10.187 & 0.034 & 9.427 & 0.073 & 5.586 & 0.221 & 1.511 & 1.315E-14\\
  CAHA\_20320053\_402026 & 20:32:00.53 & +40:20:27.0 & 15.688 & 0.003 & 13.881 & 0.005 & 12.739 & 0.005 & 11.66 & 0.017 & 10.921 & 0.016 & 10.346 & 0.027 & 9.495 & 0.049 & 5.696 & 0.133 & --- & --- \\
  CAHA\_20325540\_402150 & 20:32:55.40 & +40:21:50.7 & 15.923 & 0.005 & 13.857 & 0.007 & 12.355 & 0.007 & 10.89 & 0.016 & 10.376 & 0.016 & 9.785 & 0.016 & 9.012 & 0.018 & 5.954 & 0.056 & --- & --- \\
  CAHA\_20330082\_401800 & 20:33:00.82 & +40:18:00.2 & 13.978 & 0.007 & 12.929 & 0.012 & 12.615 & 0.026 & 11.539 & 0.016 & 11.08 & 0.016 & 10.769 & 0.018 & 9.75 & 0.022 & 6.142 & 0.09 & --- & --- \\
  CAHA\_20322395\_402218 & 20:32:23.95 & +40:22:18.3 & 15.943 & 0.037 & 14.013 & 0.035 & 12.994 & 0.033 & 11.783 & 0.016 & 11.241 & 0.016 & 10.787 & 0.021 & 10.031 & 0.034 & 6.152 & 0.158 & --- & --- \\
  CAHA\_20321650\_402141 & 20:32:16.50 & +40:21:41.1 & 18.196 & 0.012 & 15.382 & 0.016 & 13.564 & 0.021 & 12.04 & 0.017 & 11.356 & 0.016 & 10.867 & 0.029 & 10.155 & 0.057 & 6.186 & 0.177 & --- & --- \\
  CAHA\_20320880\_402054 & 20:32:08.80 & +40:20:54.9 & 15.698 & 0.003 & 13.914 & 0.006 & 12.866 & 0.004 & 11.565 & 0.016 & 10.84 & 0.016 & 10.039 & 0.025 & 8.887 & 0.039 & 6.378 & 0.098 & --- & --- \\
  CAHA\_20324730\_402217 & 20:32:47.30 & +40:22:17.9 & 14.143 & 0.004 & 13.131 & 0.006 & 12.514 & 0.006 & 11.697 & 0.016 & 11.226 & 0.016 & 10.764 & 0.019 & 10.016 & 0.021 & 6.462 & 0.123 & --- & --- \\
  CAHA\_20321184\_401733 & 20:32:11.84 & +40:17:33.8 & 14.27 & 0.002 & 12.991 & 0.005 & 12.4 & 0.009 & 11.543 & 0.017 & 11.161 & 0.017 & 10.727 & 0.055 & 10.004 & 0.086 & 6.559 & 0.163 & --- & --- \\
  CAHA\_20325680\_401223 & 20:32:56.80 & +40:12:23.3 & 15.594 & 0.015 & 13.946 & 0.017 & 12.806 & 0.017 & 11.997 & 0.016 & 11.55 & 0.016 & 11.297 & 0.021 & 10.649 & 0.029 & 6.835 & 0.132 & --- & --- \\
  CAHA\_20330112\_401449 & 20:33:01.12 & +40:14:49.2 & 14.081 & 0.011 & 13.502 & 0.036 & 12.493 & 0.025 & 11.214 & 0.016 & 10.878 & 0.016 & 10.53 & 0.018 & 9.727 & 0.02 & 6.93 & 0.237 & --- & --- \\
  CAHA\_20330982\_401154 & 20:33:09.82 & +40:11:54.6 & 14.388 & 0.01 & 13.128 & 0.012 & 12.538 & 0.024 & 11.265 & 0.016 & 10.82 & 0.016 & 10.411 & 0.017 & 9.723 & 0.017 & 6.99 & 0.199 & --- & --- \\
  CAHA\_20325220\_402406 & 20:32:52.20 & +40:24:06.2 & 15.228 & 0.006 & 13.653 & 0.008 & 12.688 & 0.009 & 11.626 & 0.016 & 11.181 & 0.016 & 10.784 & 0.018 & 10.062 & 0.029 & 7.023 & 0.064 & --- & --- \\
  CAHA\_20324795\_402214 & 20:32:47.95 & +40:22:14.2 & 18.103 & 0.007 & 15.928 & 0.007 & 14.661 & 0.007 & 13.505 & 0.038 & 12.873 & 0.032 & 12.302 & 0.053 & 11.42 & 0.07 & 7.027 & 0.246 & --- & --- \\
  CAHA\_20325444\_401949 & 20:32:54.44 & +40:19:49.5 & 14.826 & 0.004 & 13.712 & 0.007 & 13.046 & 0.005 & 12.011 & 0.016 & 11.721 & 0.016 & 11.454 & 0.025 & 11.034 & 0.037 & 7.044 & 0.183 & --- & --- \\
  CAHA\_20320189\_401007 & 20:32:01.89 & +40:10:07.8 & 16.296 & 0.009 & 14.641 & 0.011 & 13.655 & 0.049 & 12.911 & 0.02 & 12.625 & 0.02 & 12.386 & 0.049 & 11.962 & 0.109 & 7.045 & 0.217 & --- & --- \\
  CAHA\_20330546\_401215 & 20:33:05.46 & +40:12:15.7 & 14.32 & 0.036 & 13.247 & 0.038 & 12.594 & 0.055 & 11.881 & 0.016 & 11.534 & 0.016 & 11.129 & 0.021 & 10.243 & 0.035 & 7.084 & 0.19 & 1.701 & 1.217E-14\\
  CAHA\_20330798\_401915 & 20:33:07.98 & +40:19:15.5 & 13.951 & 0.007 & 12.724 & 0.01 & 12.251 & 0.024 & 11.21 & 0.016 & 10.83 & 0.016 & 10.462 & 0.017 & 9.941 & 0.022 & 7.182 & 0.209 & --- & --- \\
  CAHA\_20321078\_402356 & 20:32:10.78 & +40:23:56.0 & 16.728 & 0.005 & 14.42 & 0.011 & 12.906 & 0.005 & 11.646 & 0.016 & 10.951 & 0.016 & 10.453 & 0.017 & 9.679 & 0.019 & 7.496 & 0.154 & --- & --- \\
  CAHA\_20331260\_402247 & 20:33:12.60 & +40:22:47.2 & 17.066 & 0.011 & 15.378 & 0.096 & 14.455 & 0.096 & 13.466 & 0.024 & 13.22 & 0.021 & 12.702 & 0.059 & 11.862 & 0.087 & 7.898 & 0.246 & --- & --- \\
  CAHA\_20321598\_401023 & 20:32:15.98 & +40:10:23.5 & 14.683 & 0.013 & 13.6 & 0.014 & 12.856 & 0.033 & 12.218 & 0.016 & 11.902 & 0.016 & 11.536 & 0.023 & 11.097 & 0.04 & --- & ---  & 1.467 & 1.295E-14\\
  CAHA\_20323581\_400914 & 20:32:35.81 & +40:09:14.3 & 16.045 & 9.999 & 14.927 & 0.018 & 13.921 & 0.023 & 13.471 & 0.021 & 13.326 & 0.022 & 12.951 & 0.121 & 12.124 & 0.243 & --- & ---  & --- & --- \\
  CAHA\_20325027\_401053 & 20:32:50.27 & +40:10:53.5 & 16.775 & 0.023 & 15.134 & 0.023 & 13.962 & 0.025 & 13.741 & 0.034 & 13.334 & 0.032 & 12.821 & 0.084 & 11.766 & 0.129 & --- & ---  & --- & --- \\
  CAHA\_20320910\_401441 & 20:32:09.10 & +40:14:41.7 & 15.85 & 0.006 & 13.621 & 0.006 & 12.556 & 0.021 & 11.916 & 0.019 & 11.779 & 0.018 & 11.381 & 0.052 & 10.657 & 0.113 & --- & ---  & --- & --- \\
  CAHA\_20321210\_401240 & 20:32:12.10 & +40:12:41.0 & 15.8 & 0.01 & 14.19 & 0.009 & 13.448 & 0.03 & 12.56 & 0.018 & 12.053 & 0.017 & 11.504 & 0.035 & 10.608 & 0.052 & --- & ---  & --- & --- \\
  CAHA\_20320975\_401131 & 20:32:09.75 & +40:11:31.3 & 17.666 & 0.01 & 15.457 & 0.011 & 14.374 & 0.035 & 13.816 & 0.027 & 13.654 & 0.033 & 13.254 & 0.131 & 12.599 & 0.17 & --- & ---  & --- & --- \\
  CAHA\_20320721\_401330 & 20:32:07.21 & +40:13:30.5 & 15.47 & 0.016 & 14.657 & 0.025 & 14.144 & 0.026 & 13.503 & 0.028 & 13.251 & 0.027 & 12.661 & 0.125 & 11.52 & 0.207 & --- & ---  & --- & --- \\
  CAHA\_20320583\_401139 & 20:32:05.83 & +40:11:39.8 & 17.571 & 0.012 & 15.648 & 0.011 & 14.458 & 0.035 & 13.53 & 0.022 & 13.079 & 0.02 & 12.704 & 0.065 & 12.004 & 0.105 & --- & ---  & --- & --- \\
  CAHA\_20321307\_401250 & 20:32:13.07 & +40:12:50.0 & 16.731 & 0.009 & 15.006 & 0.01 & 13.919 & 0.027 & 12.918 & 0.024 & 12.355 & 0.021 & 12.049 & 0.051 & 11.509 & 0.141 & --- & ---  & --- & --- \\
  CAHA\_20321094\_401157 & 20:32:10.94 & +40:11:57.3 & 14.802 & 0.01 & 13.906 & 0.009 & 13.395 & 0.03 & 13.259 & 0.022 & 13.186 & 0.025 & 12.688 & 0.096 & 11.88 & 0.237 & --- & ---  & --- & --- \\
  CAHA\_20321434\_401357 & 20:32:14.34 & +40:13:57.8 & 16.7 & 0.152 & 14.895 & 0.062 & 14.25 & 0.024 & 12.878 & 0.017 & 12.516 & 0.019 & 12.323 & 0.056 & 11.361 & 0.082 & --- & ---  & --- & --- \\
  CAHA\_20320296\_401350 & 20:32:02.96 & +40:13:50.9 & 17.226 & 0.009 & 14.281 & 0.007 & 12.922 & 0.024 & 12.062 & 0.018 & 11.914 & 0.019 & 11.574 & 0.078 & 10.965 & 0.228 & --- & ---  & --- & --- \\
  CAHA\_20320488\_401316 & 20:32:04.88 & +40:13:16.0 & 16.034 & 0.007 & 14.24 & 0.007 & 13.273 & 0.026 & 12.802 & 0.022 & 12.554 & 0.022 & 11.825 & 0.082 & 10.472 & 0.128 & --- & ---  & --- & --- \\
  CAHA\_20320511\_401231 & 20:32:05.11 & +40:12:31.7 & 19.14 & 0.027 & 15.766 & 0.01 & 14.088 & 0.03 & 13.316 & 0.027 & 13.068 & 0.026 & 12.75 & 0.106 & 11.833 & 0.237 & --- & ---  & --- & --- \\
  CAHA\_20321626\_401254 & 20:32:16.26 & +40:12:54.2 & 17.864 & 0.009 & 15.707 & 0.01 & 14.282 & 0.023 & 12.711 & 0.017 & 12.039 & 0.017 & 11.581 & 0.049 & 11.111 & 0.103 & --- & ---  & --- & --- \\
  CAHA\_20323916\_401223 & 20:32:39.16 & +40:12:23.2 & 14.6 & 0.018 & 12.759 & 0.022 & 12.049 & 0.022 & 11.536 & 0.017 & 11.475 & 0.019 & 11.083 & 0.083 & 10.366 & 0.216 & --- & ---  & --- & --- \\
  CAHA\_20323223\_401347 & 20:32:32.23 & +40:13:47.1 & 17.797 & 0.01 & 15.102 & 0.01 & 13.512 & 0.012 & 11.859 & 0.046 & 11.264 & 0.046 & 9.956 & 0.147 & 8.584 & 0.184 & --- & ---  & --- & --- \\
  CAHA\_20323296\_401334 & 20:32:32.96 & +40:13:34.5 & 14.659 & 0.008 & 12.915 & 0.021 & 12.118 & 0.022 & 10.675 & 0.024 & 10.098 & 0.021 & 9.251 & 0.097 & 8.062 & 0.161 & --- & ---  & 1.920 & 1.205E-14\\
  CAHA\_20323055\_401356 & 20:32:30.55 & +40:13:56.2 & 18.796 & 0.015 & 15.275 & 0.01 & 13.144 & 0.017 & 11.272 & 0.046 & 10.48 & 0.033 & 9.671 & 0.127 & 8.665 & 0.228 & --- & ---  & 3.059 & 1.572E-15\\
  CAHA\_20323042\_401221 & 20:32:30.42 & +40:12:21.1 & 17.321 & 0.015 & 15.434 & 0.015 & 14.292 & 0.022 & 13.077 & 0.045 & 12.658 & 0.028 & 11.364 & 0.121 & 10.082 & 0.182 & --- & ---  & --- & --- \\
  CAHA\_20323549\_401308 & 20:32:35.49 & +40:13:08.9 & 17.794 & 0.012 & 14.456 & 0.012 & 12.912 & 0.023 & 11.898 & 0.024 & 11.703 & 0.023 & 10.857 & 0.111 & 9.75 & 0.194 & --- & ---  & --- & --- \\
  CAHA\_20325130\_401150 & 20:32:51.30 & +40:11:50.8 & 17.119 & 0.016 & 15.336 & 0.016 & 14.289 & 0.016 & 13.299 & 0.03 & 12.742 & 0.021 & 12.286 & 0.049 & 11.36 & 0.058 & --- & ---  & --- & --- \\
  CAHA\_20325783\_401208 & 20:32:57.83 & +40:12:08.4 & 15.826 & 0.015 & 14.792 & 0.017 & 14.297 & 0.021 & 14.107 & 0.027 & 13.942 & 0.029 & 13.696 & 0.095 & 12.871 & 0.135 & --- & ---  & --- & --- \\
  CAHA\_20330832\_401317 & 20:33:08.32 & +40:13:17.2 & 15.914 & 0.01 & 15.062 & 0.016 & 14.513 & 0.029 & 14.045 & 0.026 & 13.567 & 0.027 & 13.146 & 0.07 & 12.456 & 0.114 & --- & ---  & --- & --- \\
  CAHA\_20330731\_401428 & 20:33:07.31 & +40:14:28.4 & 15.442 & 0.009 & 14.386 & 0.016 & 13.868 & 0.021 & 13.555 & 0.03 & 13.252 & 0.029 & 12.772 & 0.049 & 12.064 & 0.09 & --- & ---  & --- & --- \\
  CAHA\_20330608\_401211 & 20:33:06.08 & +40:12:11.6 & 14.922 & 0.01 & 13.97 & 0.015 & 13.365 & 0.036 & 13.081 & 0.023 & 12.807 & 0.023 & 12.479 & 0.048 & 11.86 & 0.058 & --- & ---  & --- & --- \\
  CAHA\_20315908\_401647 & 20:31:59.08 & +40:16:47.9 & 14.435 & 0.006 & 13.58 & 0.008 & 13.221 & 0.013 & 12.687 & 0.021 & 12.142 & 0.019 & 11.856 & 0.057 & 10.977 & 0.139 & --- & ---  & 1.570 & 7.604E-15\\
  CAHA\_20320234\_401726 & 20:32:02.34 & +40:17:26.9 & 18.316 & 0.011 & 15.371 & 0.007 & 13.777 & 0.008 & 12.528 & 0.023 & 11.981 & 0.02 & 11.466 & 0.051 & 10.813 & 0.088 & --- & ---  & --- & --- \\
  CAHA\_20320067\_401621 & 20:32:00.67 & +40:16:22.0 & 14.559 & 0.004 & 13.399 & 0.006 & 12.684 & 0.014 & 12.3 & 0.024 & 12.007 & 0.023 & 11.448 & 0.053 & 10.343 & 0.062 & --- & ---  & --- & --- \\
  CAHA\_20321121\_401703 & 20:32:11.21 & +40:17:03.5 & 16.9 & 0.004 & 15.195 & 0.006 & 14.244 & 0.011 & 12.913 & 0.022 & 12.358 & 0.02 & 11.972 & 0.076 & 11.223 & 0.171 & --- & ---  & --- & --- \\
  CAHA\_20321075\_401624 & 20:32:10.75 & +40:16:24.8 & 13.638 & 0.004 & 12.659 & 0.006 & 11.911 & 0.018 & 11.199 & 0.016 & 10.843 & 0.016 & 10.509 & 0.019 & 10.005 & 0.057 & --- & ---  & --- & --- \\
  CAHA\_20320290\_401724 & 20:32:02.90 & +40:17:24.4 & 16.272 & 0.005 & 15.33 & 0.027 & 14.234 & 0.021 & 12.75 & 0.04 & 12.243 & 0.033 & 11.751 & 0.074 & 11.132 & 0.178 & --- & ---  & --- & --- \\
  CAHA\_20320046\_401506 & 20:32:00.46 & +40:15:06.1 & 15.626 & 0.006 & 13.109 & 0.009 & 12.109 & 0.018 & 11.339 & 0.023 & 11.244 & 0.023 & 10.875 & 0.043 & 10.259 & 0.107 & --- & ---  & --- & --- \\
  CAHA\_20320665\_401519 & 20:32:06.65 & +40:15:20.0 & 14.918 & 0.005 & 12.888 & 0.007 & 11.834 & 0.02 & 10.679 & 0.016 & 10.308 & 0.016 & 9.932 & 0.038 & 9.529 & 0.117 & --- & ---  & --- & --- \\
  CAHA\_20320783\_401636 & 20:32:07.83 & +40:16:36.3 & 15.209 & 0.004 & 14.078 & 0.007 & 13.417 & 0.014 & 12.87 & 0.019 & 12.576 & 0.019 & 12.233 & 0.058 & 11.23 & 0.176 & --- & ---  & --- & --- \\
  CAHA\_20320403\_401828 & 20:32:04.03 & +40:18:28.1 & 20.924 & 0.058 & 17.029 & 0.012 & 14.745 & 0.008 & 13.107 & 0.033 & 12.324 & 0.023 & 11.886 & 0.098 & 10.963 & 0.229 & --- & ---  & --- & --- \\
  CAHA\_20320309\_401738 & 20:32:03.09 & +40:17:38.4 & 19.014 & 0.018 & 15.72 & 0.007 & 13.95 & 0.009 & 12.491 & 0.021 & 11.803 & 0.02 & 11.231 & 0.048 & 10.465 & 0.089 & --- & ---  & --- & --- \\
  CAHA\_20320807\_401715 & 20:32:08.07 & +40:17:15.7 & 14.595 & 0.003 & 13.256 & 0.006 & 12.477 & 0.014 & 11.741 & 0.016 & 11.405 & 0.016 & 11.075 & 0.033 & 10.171 & 0.121 & --- & ---  & --- & --- \\
  CAHA\_20320848\_401550 & 20:32:08.48 & +40:15:50.7 & 16.506 & 0.003 & 15.376 & 0.006 & 14.944 & 0.015 & 14.359 & 0.061 & 14.224 & 0.063 & 12.389 & 0.102 & 10.463 & 0.09 & --- & ---  & --- & --- \\
  CAHA\_20322058\_401825 & 20:32:20.58 & +40:18:25.4 & 18.092 & 0.007 & 15.837 & 0.012 & 14.668 & 0.023 & 12.748 & 0.023 & 12.0 & 0.02 & 11.443 & 0.074 & 10.774 & 0.114 & --- & ---  & --- & --- \\
  CAHA\_20322443\_401818 & 20:32:24.43 & +40:18:18.4 & 16.859 & 0.024 & 14.573 & 0.024 & 13.207 & 0.024 & 11.911 & 0.021 & 11.259 & 0.02 & 10.694 & 0.083 & 9.749 & 0.196 & --- & ---  & 2.913 & 1.306E-15\\
  CAHA\_20322935\_401844 & 20:32:29.35 & +40:18:44.7 & 14.703 & 0.003 & 13.689 & 0.005 & 13.172 & 0.004 & 12.806 & 0.021 & 12.618 & 0.024 & 12.336 & 0.124 & 11.746 & 0.24 & --- & ---  & --- & --- \\
  CAHA\_20321655\_401808 & 20:32:16.55 & +40:18:08.8 & 15.139 & 0.003 & 13.688 & 0.007 & 12.914 & 0.005 & 12.245 & 0.022 & 11.862 & 0.019 & 11.021 & 0.072 & 9.779 & 0.069 & --- & ---  & --- & --- \\
  CAHA\_20323555\_401840 & 20:32:35.55 & +40:18:40.3 & 18.715 & 0.015 & 15.218 & 0.005 & 13.545 & 0.006 & 12.638 & 0.022 & 12.426 & 0.023 & 12.034 & 0.112 & 11.721 & 0.239 & --- & ---  & --- & --- \\
  CAHA\_20323489\_401810 & 20:32:34.89 & +40:18:11.0 & 13.619 & 0.004 & 12.464 & 0.005 & 11.896 & 0.021 & 10.95 & 0.016 & 10.578 & 0.016 & 10.286 & 0.043 & 9.551 & 0.082 & --- & ---  & 1.219 & 1.186E-15\\
  CAHA\_20323249\_401602 & 20:32:32.49 & +40:16:02.9 & 17.053 & 0.027 & 13.156 & 0.007 & 10.815 & 0.017 & 8.458 & 0.017 & 7.656 & 0.017 & 7.065 & 0.034 & 6.613 & 0.093 & --- & ---  & --- & --- \\
  CAHA\_20323176\_401616 & 20:32:31.76 & +40:16:16.5 & 14.715 & 0.005 & 11.98 & 0.018 & 10.456 & 0.017 & 9.25 & 0.031 & 8.842 & 0.054 & 7.789 & 0.142 & 5.961 & 0.131 & --- & ---  & --- & --- \\
  CAHA\_20324131\_401807 & 20:32:41.31 & +40:18:07.2 & 17.244 & 0.005 & 15.14 & 0.007 & 14.169 & 0.007 & 13.383 & 0.021 & 12.98 & 0.022 & 12.605 & 0.101 & 11.901 & 0.207 & --- & ---  & --- & --- \\
  CAHA\_20323108\_401608 & 20:32:31.08 & +40:16:08.6 & 16.236 & 0.009 & 13.524 & 0.005 & 12.042 & 0.032 & 10.022 & 0.044 & 9.314 & 0.046 & 8.55 & 0.133 & 6.872 & 0.167 & --- & ---  & --- & --- \\
  CAHA\_20324014\_401812 & 20:32:40.14 & +40:18:12.7 & 16.204 & 0.004 & 14.726 & 0.007 & 13.956 & 0.005 & 13.485 & 0.029 & 13.178 & 0.035 & 12.759 & 0.13 & 11.857 & 0.213 & --- & ---  & --- & --- \\
  CAHA\_20323095\_401649 & 20:32:30.95 & +40:16:49.6 & 13.021 & 0.004 & 10.756 & 0.022 & 9.763 & 0.017 & 9.074 & 0.032 & 8.531 & 0.031 & 7.958 & 0.127 & 6.609 & 0.208 & --- & ---  & 2.708 & 6.967E-15\\
  CAHA\_20323708\_401737 & 20:32:37.08 & +40:17:37.5 & 13.577 & 0.004 & 11.519 & 0.031 & 10.529 & 0.026 & 9.853 & 0.023 & 9.483 & 0.027 & 8.839 & 0.084 & 7.252 & 0.093 & --- & ---  & 2.110 & 3.729E-15\\
  CAHA\_20323555\_401605 & 20:32:35.55 & +40:16:05.8 & 13.195 & 0.007 & 10.399 & 0.022 & 8.493 & 0.017 & 6.322 & 0.016 & 5.666 & 0.015 & 5.036 & 0.021 & 4.051 & 0.038 & --- & ---  & --- & --- \\
  CAHA\_20330586\_401552 & 20:33:05.86 & +40:15:52.9 & 17.201 & 0.014 & 15.822 & 0.028 & 15.098 & 0.043 & 14.265 & 0.027 & 13.921 & 0.024 & 13.669 & 0.099 & 13.24 & 0.204 & --- & ---  & --- & --- \\
  CAHA\_20320906\_402043 & 20:32:09.06 & +40:20:43.0 & 14.099 & 0.002 & 12.956 & 0.006 & 12.338 & 0.004 & 12.017 & 0.017 & 11.927 & 0.018 & 11.576 & 0.078 & 10.78 & 0.15 & --- & ---  & --- & --- \\
  CAHA\_20321381\_401908 & 20:32:13.81 & +40:19:08.6 & 20.069 & 0.057 & 16.556 & 0.007 & 14.37 & 0.006 & 12.346 & 0.092 & 11.718 & 0.054 & 10.147 & 0.178 & 8.759 & 0.227 & --- & ---  & --- & --- \\
  CAHA\_20322602\_401904 & 20:32:26.02 & +40:19:04.3 & 18.818 & 9.999 & 16.463 & 0.008 & 14.207 & 0.005 & 12.653 & 0.017 & 12.093 & 0.017 & 11.66 & 0.049 & 11.05 & 0.116 & --- & ---  & --- & --- \\
  CAHA\_20322626\_402216 & 20:32:26.26 & +40:22:16.8 & 15.245 & 0.003 & 13.693 & 0.007 & 12.927 & 0.004 & 12.298 & 0.017 & 11.832 & 0.016 & 11.483 & 0.035 & 10.776 & 0.072 & --- & ---  & --- & --- \\
  CAHA\_20322761\_401914 & 20:32:27.61 & +40:19:14.5 & 20.543 & 0.057 & 16.125 & 0.006 & 13.617 & 0.005 & 12.102 & 0.017 & 11.45 & 0.017 & 11.021 & 0.045 & 10.301 & 0.058 & --- & ---  & --- & --- \\
  CAHA\_20322942\_401917 & 20:32:29.42 & +40:19:17.2 & 19.044 & 0.015 & 16.167 & 0.005 & 14.449 & 0.009 & 13.061 & 0.02 & 12.395 & 0.021 & 12.073 & 0.054 & 11.34 & 0.107 & --- & ---  & --- & --- \\
  CAHA\_20322974\_401906 & 20:32:29.74 & +40:19:06.3 & 18.843 & 0.012 & 16.098 & 0.007 & 14.651 & 0.007 & 13.529 & 0.025 & 12.964 & 0.023 & 12.437 & 0.067 & 11.733 & 0.081 & --- & ---  & --- & --- \\
  CAHA\_20322536\_402054 & 20:32:25.36 & +40:20:54.5 & 14.601 & 0.003 & 12.966 & 0.006 & 12.168 & 0.004 & 11.392 & 0.017 & 11.041 & 0.016 & 10.586 & 0.019 & 9.925 & 0.055 & --- & ---  & --- & --- \\
  CAHA\_20322352\_402035 & 20:32:23.52 & +40:20:35.2 & 19.808 & 0.038 & 16.256 & 0.007 & 14.394 & 0.007 & 13.064 & 0.023 & 12.391 & 0.02 & 11.636 & 0.059 & 10.866 & 0.08 & --- & ---  & --- & --- \\
  CAHA\_20322337\_402212 & 20:32:23.37 & +40:22:12.2 & 15.341 & 0.003 & 13.874 & 0.008 & 13.182 & 0.004 & 12.554 & 0.017 & 12.186 & 0.017 & 11.763 & 0.028 & 10.759 & 0.04 & --- & ---  & --- & --- \\
  CAHA\_20321759\_402226 & 20:32:17.59 & +40:22:26.4 & 17.119 & 0.006 & 15.531 & 0.008 & 14.738 & 0.006 & 13.952 & 0.028 & 13.531 & 0.024 & 13.111 & 0.081 & 12.349 & 0.072 & --- & ---  & --- & --- \\
  CAHA\_20321778\_401905 & 20:32:17.78 & +40:19:05.8 & 15.883 & 0.003 & 13.723 & 0.006 & 12.612 & 0.005 & 11.801 & 0.035 & 11.492 & 0.029 & 10.374 & 0.156 & 9.075 & 0.23 & --- & ---  & --- & --- \\
  CAHA\_20321874\_402003 & 20:32:18.74 & +40:20:03.2 & 19.097 & 0.014 & 16.012 & 0.007 & 14.24 & 0.007 & 12.801 & 0.018 & 12.146 & 0.023 & 11.508 & 0.063 & 10.417 & 0.126 & --- & ---  & --- & --- \\
  CAHA\_20321559\_402131 & 20:32:15.59 & +40:21:31.2 & 17.573 & 0.005 & 15.115 & 0.007 & 13.785 & 0.004 & 13.103 & 0.019 & 12.853 & 0.02 & 12.689 & 0.113 & 11.855 & 0.246 & --- & ---  & --- & --- \\
  CAHA\_20323947\_402116 & 20:32:39.47 & +40:21:16.5 & 15.868 & 0.003 & 14.169 & 0.006 & 13.308 & 0.005 & 12.546 & 0.018 & 12.045 & 0.017 & 11.555 & 0.034 & 10.922 & 0.063 & --- & ---  & --- & --- \\
  CAHA\_20324423\_401939 & 20:32:44.23 & +40:19:39.4 & 16.834 & 0.004 & 15.494 & 0.007 & 14.667 & 0.009 & 13.761 & 0.023 & 13.168 & 0.022 & 12.481 & 0.054 & 11.586 & 0.071 & --- & ---  & --- & --- \\
  CAHA\_20323153\_402048 & 20:32:31.53 & +40:20:49.0 & 18.767 & 0.018 & 15.671 & 0.006 & 14.18 & 0.009 & 13.323 & 0.02 & 13.097 & 0.021 & 12.761 & 0.058 & 12.403 & 0.122 & --- & ---  & --- & --- \\
  CAHA\_20323570\_402141 & 20:32:35.70 & +40:21:41.9 & 18.206 & 0.012 & 15.383 & 0.006 & 14.069 & 0.006 & 13.276 & 0.019 & 13.112 & 0.02 & 12.83 & 0.071 & 12.158 & 0.155 & --- & ---  & --- & --- \\
  CAHA\_20323809\_401943 & 20:32:38.09 & +40:19:43.7 & 18.638 & 0.015 & 15.449 & 0.006 & 13.945 & 0.007 & 13.14 & 0.02 & 12.916 & 0.02 & 12.593 & 0.078 & 12.069 & 0.157 & --- & ---  & --- & --- \\
  CAHA\_20323774\_401916 & 20:32:37.74 & +40:19:16.4 & 20.18 & 0.044 & 16.788 & 0.011 & 14.861 & 0.006 & 13.415 & 0.021 & 12.829 & 0.021 & 12.384 & 0.095 & 11.399 & 0.107 & --- & ---  & --- & --- \\
  CAHA\_20325212\_401914 & 20:32:52.12 & +40:19:14.8 & 17.619 & 0.004 & 15.916 & 0.008 & 15.151 & 0.012 & 14.45 & 0.043 & 13.974 & 0.038 & 13.495 & 0.118 & 12.946 & 0.175 & --- & ---  & --- & --- \\
  CAHA\_20325635\_402129 & 20:32:56.35 & +40:21:29.0 & 17.837 & 0.008 & 15.611 & 0.007 & 14.432 & 0.009 & 13.374 & 0.019 & 12.82 & 0.017 & 12.432 & 0.035 & 11.756 & 0.056 & --- & ---  & --- & --- \\
  CAHA\_20315935\_402411 & 20:31:59.35 & +40:24:11.4 & 15.12 & 0.006 & 13.917 & 0.012 & 13.485 & 0.043 & 12.841 & 0.018 & 12.476 & 0.017 & 12.116 & 0.032 & 11.572 & 0.049 & --- & ---  & --- & --- \\
  CAHA\_20315954\_402231 & 20:31:59.54 & +40:22:31.8 & 16.638 & 0.004 & 15.201 & 0.008 & 14.36 & 0.006 & 13.77 & 0.024 & 13.34 & 0.024 & 13.058 & 0.063 & 12.413 & 0.135 & --- & ---  & --- & --- \\
  CAHA\_20320915\_402301 & 20:32:09.15 & +40:23:01.9 & 16.366 & 0.004 & 14.708 & 0.009 & 13.754 & 0.004 & 12.628 & 0.022 & 12.111 & 0.019 & 11.678 & 0.047 & 10.876 & 0.095 & --- & ---  & --- & --- \\
  CAHA\_20320939\_402250 & 20:32:09.39 & +40:22:50.6 & 13.076 & 0.004 & 11.106 & 0.023 & 10.307 & 0.017 & 9.56 & 0.016 & 9.078 & 0.016 & 8.597 & 0.016 & 7.961 & 0.017 & --- & ---  & 1.920 & 3.140E-14\\
  CAHA\_20320112\_402313 & 20:32:01.12 & +40:23:13.2 & 17.417 & 0.019 & 15.762 & 0.03 & 14.796 & 0.039 & 13.528 & 0.019 & 12.909 & 0.017 & 12.462 & 0.047 & 11.808 & 0.077 & --- & ---  & --- & --- \\
  CAHA\_20322465\_402329 & 20:32:24.65 & +40:23:29.2 & 17.824 & 0.008 & 15.397 & 0.009 & 14.258 & 0.006 & 13.65 & 0.023 & 13.479 & 0.023 & 13.182 & 0.062 & 12.632 & 0.166 & --- & ---  & --- & --- \\
  CAHA\_20324555\_402331 & 20:32:45.55 & +40:23:31.7 & 17.385 & 0.006 & 15.647 & 0.007 & 14.723 & 0.005 & 14.097 & 0.024 & 13.687 & 0.023 & 13.609 & 0.115 & 12.844 & 0.202 & --- & ---  & --- & --- \\
  CAHA\_20324550\_402231 & 20:32:45.50 & +40:22:31.7 & 16.875 & 0.004 & 15.837 & 0.007 & 15.174 & 0.005 & 14.419 & 0.038 & 13.977 & 0.033 & 13.046 & 0.059 & 12.014 & 0.091 & --- & ---  & --- & --- \\
  CAHA\_20330316\_402332 & 20:33:03.16 & +40:23:33.0 & 18.123 & 0.031 & 15.717 & 0.026 & 14.347 & 0.018 & 12.893 & 0.019 & 12.421 & 0.017 & 12.128 & 0.046 & 11.701 & 0.117 & --- & ---  & --- & --- \\
  TWOM\_20324634\_4009048 & 20:32:46.35 & +40:09:04.9 & 18.49 & 9.999 & 14.906 & 0.059 & 13.206 & 0.036 & 12.13 & 0.017 & 11.948 & 0.017 & 11.546 & 0.038 & 11.182 & 0.062 & --- & ---  & --- & --- \\
  TWOM\_20315282\_4012188 & 20:31:52.82 & +40:12:18.8 & 13.994 & 0.027 & 13.068 & 0.029 & 12.612 & 0.026 & 12.05 & 0.017 & 11.775 & 0.017 & 11.533 & 0.027 & 11.024 & 0.046 & --- & ---  & --- & --- \\
  TWOM\_20322814\_4017148 & 20:32:28.14 & +40:17:14.9 & 10.966 & 0.035 & 9.926 & 0.038 & 9.336 & 0.024 & 8.985 & 0.034 & 8.541 & 0.046 & 7.94 & 0.172 & 6.448 & 0.242 & --- & ---  & --- & --- \\
  TWOM\_20331278\_4018418 & 20:33:12.79 & +40:18:41.8 & 18.61 & 9.999 & 15.771 & 0.13 & 14.561 & 0.103 & 13.589 & 0.029 & 13.368 & 0.024 & 13.103 & 0.098 & 12.477 & 0.185 & --- & ---  & --- & --- \\
  CAHA\_20325255\_401152 & 20:32:52.55 & +40:11:52.0 & 19.187 & 0.034 & 17.342 & 0.018 & 16.09 & 0.017 & 14.485 & 0.027 & 13.904 & 0.024 & 13.684 & 0.111 & 12.616 & 0.157 & --- & ---  & --- & --- \\
  CAHA\_20321844\_401724 & 20:32:18.23 & +40:17:27.7 & 18.294 & 0.033 & 16.583 & 0.015 & 15.157 & 0.007 & 13.024 & 0.026 & 12.173 & 0.019 & 11.537 & 0.046 & 10.547 & 0.045 & --- & ---  & --- & --- \\
  CAHA\_20322275\_401739 & 20:32:22.75 & +40:17:39.6 & 20.249 & 0.058 & 18.616 & 0.064 & 17.053 & 0.089 & 12.84 & 0.153 & 12.698 & 0.187 & 9.895 & 0.171 & 8.093 & 0.184 & --- & ---  & --- & --- \\
  CAHA\_20324785\_401817 & 20:32:47.85 & +40:18:17.7 & 20.327 & 0.059 & 16.924 & 0.008 & 15.097 & 0.008 & 14.101 & 0.04 & 13.779 & 0.046 & 13.599 & 0.16 & 12.532 & 0.219 & --- & ---  & --- & --- \\
  CAHA\_20324105\_401849 & 20:32:41.05 & +40:18:49.3 & 18.836 & 0.017 & 16.75 & 0.007 & 15.68 & 0.012 & 14.572 & 0.041 & 14.287 & 0.038 & 13.87 & 0.173 & 12.149 & 0.14 & --- & ---  & --- & --- \\
  CAHA\_20325040\_401850 & 20:32:50.40 & +40:18:50.4 & 17.912 & 0.009 & 16.288 & 0.01 & 15.384 & 0.01 & 14.69 & 0.056 & 14.384 & 0.051 & 13.923 & 0.118 & 13.129 & 0.199 & --- & ---  & --- & --- \\
  CAHA\_20322563\_401850 & 20:32:25.63 & +40:18:50.8 & 20.371 & 0.064 & 17.687 & 0.01 & 16.093 & 0.02 & 14.339 & 0.037 & 13.47 & 0.029 & 13.026 & 0.19 & 11.419 & 0.218 & --- & ---  & --- & --- \\
  CAHA\_20330273\_401903 & 20:33:02.73 & +40:19:03.2 & 18.929 & 0.019 & 16.783 & 0.015 & 15.583 & 0.011 & 14.568 & 0.054 & 14.16 & 0.054 & 13.824 & 0.121 & 12.891 & 0.146 & --- & ---  & --- & --- \\
  CAHA\_20323802\_401934 & 20:32:38.02 & +40:19:34.4 & 20.414 & 0.051 & 17.149 & 0.012 & 15.2 & 0.009 & 13.652 & 0.023 & 13.145 & 0.021 & 12.602 & 0.075 & 11.952 & 0.115 & --- & ---  & --- & --- \\
  CAHA\_20320973\_402253 & 20:32:09.73 & +40:22:53.4 & 13.802 & 0.004 & 12.505 & 0.009 & --- & ---  & 10.709 & 0.017 & 10.351 & 0.017 & 9.975 & 0.02 & 9.154 & 0.025 & --- & ---  & --- & --- \\
  CAHA\_20315905\_401755 & 20:31:59.05 & +40:17:56.0 & --- & ---  & 18.606 & 0.03 & 16.205 & 0.017 & 14.226 & 0.057 & 13.447 & 0.055 & 12.447 & 0.145 & 11.06 & 0.236 & --- & ---  & --- & --- \\
  CAHA\_20322723\_401923 & 20:32:27.23 & +40:19:23.1 & --- & ---  & 17.445 & 0.011 & 14.791 & 0.008 & 12.988 & 0.029 & 12.078 & 0.03 & 11.483 & 0.095 & 10.825 & 0.189 & --- & ---  & --- & --- \\
  CAHA\_20321904\_401942 & 20:32:19.04 & +40:19:42.4 & --- & ---  & 17.846 & 0.011 & 14.913 & 0.009 & 12.756 & 0.017 & 12.021 & 0.018 & 11.126 & 0.068 & 9.878 & 0.125 & --- & ---  & --- & --- \\
  CAHA\_20321823\_402039 & 20:32:18.23 & +40:20:39.4 & --- & ---  & 19.331 & 0.053 & 15.653 & 0.02 & 12.337 & 0.017 & 11.325 & 0.016 & 10.716 & 0.061 & 10.249 & 0.167 & --- & ---  & --- & --- \\
  CAHA\_20321538\_402356 & 20:32:15.38 & +40:23:56.9 & --- & ---  & 17.495 & 0.023 & 15.832 & 0.034 & 14.414 & 0.024 & 13.942 & 0.025 & 13.44 & 0.08 & 12.978 & 0.241 & --- & ---  & --- & --- \\
  CAHA\_20330081\_401020 & 20:33:00.81 & +40:10:20.1 & --- & ---  & --- & ---  & 18.342 & 0.121 & 15.457 & 0.055 & 14.702 & 0.061 & 13.348 & 0.082 & 12.192 & 0.091 & --- & ---  & --- & --- \\
  CAHA\_20322687\_401910 & 20:32:26.87 & +40:19:10.3 & --- & ---  & --- & ---  & 12.997 & 0.006 & 11.497 & 0.016 & 10.941 & 0.016 & 10.605 & 0.031 & 10.164 & 0.069 & --- & ---  & 1.832 & 3.097E-15\\
  CAHA\_20322007\_401933 & 20:32:20.07 & +40:19:33.6 & --- & ---  & --- & ---  & 16.563 & 0.02 & 13.016 & 0.041 & 11.857 & 0.027 & 11.221 & 0.13 & 9.785 & 0.223 & --- & ---  & --- & --- \\
  SSTCYGX\_J203152.69\_401840.5 & 20:31:52.68 & +40:18:40.5 & --- & ---  & --- & ---  & --- & ---  & 13.805 & 0.022 & 12.985 & 0.021 & 12.558 & 0.048 & 12.053 & 0.175 & --- & ---  & --- & --- \\
  CAHA\_20321015\_401847 & 20:32:10.15 & +40:18:48.0 & --- & ---  & 17.561 & 0.014 & 14.716 & 0.008 & 13.155 & 0.074 & 12.549 & 0.065 & 11.844 & 0.44 & 10.831 & 0.807 & --- & ---  & 2.431 & 2.387E-15\\
  CAHA\_20321798\_401922 & 20:32:17.98 & +40:19:22.3 & 19.178 & 0.018 & 15.984 & 0.006 & 14.341 & 0.006 & 12.932 & 0.077 & 12.54 & 0.056 & 11.309 & 0.362 & --- & ---  & --- & ---  & 3.409 & 1.067E-14\\
  CAHA\_20321061\_401904 & 20:32:10.61 & +40:19:04.8 & --- & ---  & --- & ---  & 17.616 & 0.05 & 14.017 & 0.067 & 13.097 & 0.081 & 11.677 & 0.173 & --- & ---  & --- & ---  & 3.176 & 1.357E-14\\
  CAHA\_20315867\_401915 & 20:31:58.67 & +40:19:15.1 & 17.58 & 0.006 & 14.666 & 0.005 & 12.948 & 0.008 & 11.306 & 0.016 & 10.714 & 0.016 & 10.313 & 0.019 & 9.896 & 0.045 & 7.262 & 0.41 & 1.862 & 9.616E-15\\
  CAHA\_20320503\_401715 & 20:32:05.03 & +40:17:16.0 & 12.833 & 0.005 & 11.862 & 0.019 & 11.351 & 0.016 & 10.564 & 0.016 & 10.174 & 0.015 & 9.762 & 0.02 & 8.992 & 0.046 & 6.837 & 0.749 & 1.380 & 6.313E-15\\
  CAHA\_20320542\_402126 & 20:32:05.42 & +40:21:26.2 & 14.416 & 0.002 & 13.22 & 0.006 & 12.679 & 0.005 & 12.127 & 0.023 & 11.936 & 0.021 & 11.7 & 0.133 & 11.008 & 0.356 & 6.932 & 0.505 & 1.920 & 1.509E-14\\
  CAHA\_20322109\_401848 & 20:32:21.09 & +40:18:48.1 & 17.17 & 0.005 & 14.563 & 0.005 & 13.189 & 0.004 & 11.819 & 0.031 & 11.28 & 0.023 & 10.886 & 0.187 & 10.03 & 0.32 & --- & ---  & 2.314 & 2.492E-15\\
  CAHA\_20322641\_401515 & 20:32:26.41 & +40:15:15.6 & 15.269 & 0.007 & 12.963 & 0.007 & 11.691 & 0.018 & 10.435 & 0.025 & 10.028 & 0.025 & 9.808 & 0.219 & 9.412 & 0.829 & --- & ---  & 1.920 & 1.220E-15\\
  CAHA\_20323239\_401643 & 20:32:32.39 & +40:16:43.3 & 15.99 & 0.005 & 13.31 & 0.005 & 11.504 & 0.026 & 10.049 & 0.071 & 9.374 & 0.043 & 9.051 & 0.477 & --- & ---  & --- & ---  & 3.059 & 7.560E-15\\
  CAHA\_20323582\_401745 & 20:32:35.82 & +40:17:45.0 & 13.798 & 0.004 & 12.611 & 0.008 & 12.108 & 0.033 & 11.136 & 0.049 & 10.972 & 0.043 & 9.44 & 0.243 & --- & ---  & --- & ---  & 2.781 & 6.038E-15\\
  CAHA\_20324487\_401834 & 20:32:44.87 & +40:18:34.1 & 15.417 & 0.005 & 13.309 & 0.008 & 12.165 & 0.021 & 11.274 & 0.016 & 10.967 & 0.016 & 10.707 & 0.022 & 10.187 & 0.048 & 7.054 & 0.322 & 2.256 & 8.162E-15\\
  CAHA\_20325504\_401617 & 20:32:55.04 & +40:16:17.3 & 13.48 & 0.009 & 12.375 & 0.021 & 11.68 & 0.019 & 10.542 & 0.015 & 10.282 & 0.015 & 10.11 & 0.017 & 9.77 & 0.017 & 6.073 & 0.163 & 1.526 & 1.505E-14\\
  CAHA\_20330681\_401337 & 20:33:06.81 & +40:13:37.9 & 13.464 & 0.017 & 12.831 & 0.024 & 12.525 & 0.027 & 11.991 & 0.016 & 11.836 & 0.017 & 11.689 & 0.027 & 11.2 & 0.044 & 7.413 & 0.258 & 1.613 & 1.026E-14\\
  
  \enddata
\label{tab:ClassII}
\end{deluxetable}

\clearpage

\begin{deluxetable}{lllllllllllllllllllll}
\tabletypesize{\tiny}
\setlength{\tabcolsep}{0.015in}
\tablecolumns{21}
\tablewidth{0pt}
\tablecaption{Class III sources in the DR15 regionn}
\tablehead{
\colhead{Source}  & 
\colhead{RA} &
\colhead{DEC} &
\colhead{J} &
\colhead{$\sigma_J$} &
\colhead{H} &
\colhead{$\sigma_H$} &
\colhead{K} &
\colhead{$\sigma_K$} &
\colhead{[3.6]} &
\colhead{$\sigma_{[3.6]}$} &
\colhead{[4.5]} &
\colhead{$\sigma_{[4.5]}$} &
\colhead{[5.8]} &
\colhead{$\sigma_{[5.8]}$} &
\colhead{[8.0]} &
\colhead{$\sigma_{[8.0]}$} &
\colhead{[24\ $\mu$m]} &
\colhead{$\sigma_{24\mu m}$} &
\colhead{Median Energy} &
\colhead{Energy Flux} \\
\colhead{} &
\multicolumn{2}{c}{[J2000]} &
\multicolumn{16}{c}{[mag]} &
\colhead{[keV]} &
\colhead{[erg/cm$^2$/s]} \\
}
\startdata

  CAHA\_20322152\_401104 & 20:32:21.52 & +40:11:04.6 & 13.842 & 0.016 & 13.002 & 0.015 & 12.596 & 0.028 & 12.559 & 0.019 & 12.548 & 0.021 & 12.395 & 0.078 & 11.608 & 0.169 & --- & ---  & 2.548 & 6.699E-15\\
  CAHA\_20323285\_401056 & 20:32:32.85 & +40:10:56.5 & 16.605 & 0.017 & 15.081 & 0.012 & 14.321 & 0.02 & 14.065 & 0.232 & 14.028 & 0.171 & --- & ---  & --- & ---  & --- & ---  & 3.438 & 6.162E-15\\
  CAHA\_20320491\_401359 & 20:32:04.91 & +40:13:59.3 & 13.227 & 0.008 & 12.717 & 0.023 & 12.476 & 0.024 & 12.571 & 0.021 & 12.592 & 0.021 & 12.509 & 0.099 & --- & ---  & --- & ---  & 1.190 & 5.738E-15\\
  CAHA\_20321916\_401318 & 20:32:19.16 & +40:13:18.2 & 15.919 & 0.011 & 14.424 & 0.01 & 13.758 & 0.025 & 13.357 & 0.055 & 13.247 & 0.067 & 13.21 & 0.55 & --- & ---  & --- & ---  & 2.489 & 3.258E-15\\
  CAHA\_20323784\_401352 & 20:32:37.84 & +40:13:52.7 & 15.577 & 0.014 & 14.011 & 0.027 & 13.419 & 0.021 & 13.344 & 0.233 & 13.284 & 0.25 & --- & ---  & --- & ---  & --- & ---  & 3.263 & 1.022E-14\\
  CAHA\_20324208\_401220 & 20:32:42.08 & +40:12:20.3 & 14.24 & 0.014 & 13.071 & 0.017 & 12.52 & 0.022 & 12.381 & 0.019 & 12.255 & 0.02 & 12.494 & 0.123 & --- & ---  & --- & ---  & 1.905 & 7.653E-15\\
  CAHA\_20323075\_401419 & 20:32:30.75 & +40:14:19.5 & 15.255 & 0.028 & 12.881 & 0.044 & 11.66 & 0.025 & 10.803 & 0.072 & 10.597 & 0.071 & --- & ---  & --- & ---  & --- & ---  & 2.694 & 1.592E-14\\
  CAHA\_20325387\_401420 & 20:32:53.87 & +40:14:20.4 & 13.2 & 0.015 & 12.069 & 0.019 & 11.614 & 0.018 & 11.383 & 0.016 & 11.305 & 0.016 & 11.215 & 0.024 & 11.149 & 0.149 & --- & ---  & 1.818 & 2.580E-14\\
  CAHA\_20325533\_401440 & 20:32:55.33 & +40:14:40.3 & 15.64 & 0.014 & 14.432 & 0.015 & 13.914 & 0.014 & 13.712 & 0.023 & 13.654 & 0.027 & 13.552 & 0.093 & --- & ---  & --- & ---  & 2.212 & 8.545E-15\\
  CAHA\_20320256\_401740 & 20:32:02.56 & +40:17:40.5 & 13.948 & 0.005 & 13.094 & 0.006 & 12.814 & 0.01 & 12.755 & 0.029 & 12.633 & 0.034 & 12.802 & 0.247 & --- & ---  & --- & ---  & 1.467 & 3.822E-15\\
  CAHA\_20320237\_401701 & 20:32:02.37 & +40:17:01.1 & 13.188 & 0.003 & 12.005 & 0.039 & 11.556 & 0.028 & 11.155 & 0.016 & 11.008 & 0.016 & 10.808 & 0.034 & 10.338 & 0.081 & --- & ---  & 1.511 & 2.660E-14\\
  CAHA\_20321032\_401752 & 20:32:10.32 & +40:17:53.0 & 14.269 & 0.003 & 13.309 & 0.006 & 12.949 & 0.009 & 12.646 & 0.03 & 12.564 & 0.043 & 12.658 & 0.378 & 12.546 & 1.721 & --- & ---  & 1.073 & 3.459E-15\\
  CAHA\_20321962\_401812 & 20:32:19.62 & +40:18:12.4 & 13.391 & 0.004 & 12.21 & 0.027 & 11.83 & 0.021 & 11.669 & 0.017 & 11.53 & 0.017 & 11.415 & 0.064 & 11.042 & 0.159 & --- & ---  & 1.292 & 3.995E-15\\
  CAHA\_20322560\_401805 & 20:32:25.60 & +40:18:05.3 & 19.153 & 0.02 & 15.839 & 0.005 & 14.207 & 0.004 & 13.319 & 0.063 & 13.104 & 0.09 & 13.092 & 0.671 & --- & ---  & --- & ---  & 3.570 & 2.168E-13\\
  CAHA\_20323482\_401818 & 20:32:34.82 & +40:18:18.4 & 14.571 & 0.021 & 13.428 & 0.028 & 12.995 & 0.029 & 12.655 & 0.027 & 12.56 & 0.037 & 12.213 & 0.136 & --- & ---  & --- & ---  & 1.438 & 4.581E-15\\
  CAHA\_20324203\_401510 & 20:32:42.03 & +40:15:10.2 & 14.755 & 0.009 & 13.799 & 0.011 & 13.423 & 0.013 & 13.169 & 0.111 & 13.281 & 0.12 & --- & ---  & --- & ---  & --- & ---  & 1.175 & 3.020E-15\\
  CAHA\_20323083\_401706 & 20:32:30.83 & +40:17:06.1 & 15.738 & 0.006 & 13.446 & 0.007 & 12.33 & 0.009 & 11.896 & 0.078 & 11.932 & 0.113 & --- & ---  & --- & ---  & --- & ---  & 3.380 & 5.798E-14\\
  CAHA\_20323043\_401643 & 20:32:30.43 & +40:16:43.2 & 15.525 & 0.003 & 13.049 & 0.006 & 11.933 & 0.026 & 11.182 & 0.055 & 11.031 & 0.106 & --- & ---  & --- & ---  & --- & ---  & 2.716 & 7.643E-15\\
  CAHA\_20325412\_401547 & 20:32:54.12 & +40:15:47.9 & 17.073 & 0.009 & 15.642 & 0.013 & 15.02 & 0.012 & 14.742 & 0.041 & 14.508 & 0.052 & --- & ---  & --- & ---  & --- & ---  & 1.978 & 1.2521E-14\\
  CAHA\_20325095\_401520 & 20:32:50.95 & +40:15:20.2 & 14.535 & 0.01 & 13.575 & 0.013 & 13.178 & 0.008 & 12.989 & 0.022 & 12.938 & 0.028 & 12.74 & 0.112 & --- & ---  & --- & ---  & 1.219 & 8.092E-15\\
  CAHA\_20315441\_402153 & 20:31:54.41 & +40:21:53.2 & 13.824 & 0.005 & 12.53 & 0.009 & 12.024 & 0.018 & 11.721 & 0.016 & 11.667 & 0.016 & 11.605 & 0.027 & 11.665 & 0.093 & --- & ---  & 1.745 & 8.670E-15\\
  CAHA\_20315532\_402127 & 20:31:55.32 & +40:21:27.6 & 14.938 & 0.002 & 13.572 & 0.008 & 12.999 & 0.005 & 12.716 & 0.02 & 12.601 & 0.021 & 12.687 & 0.105 & 12.069 & 0.322 & --- & ---  & 1.657 & 3.869E-15\\
  CAHA\_20320065\_401931 & 20:32:00.65 & +40:19:31.8 & 17.586 & 0.005 & 15.52 & 0.006 & 14.493 & 0.008 & 13.831 & 0.039 & 13.648 & 0.036 & 13.508 & 0.404 & --- & ---  & --- & ---  & 2.022 & 5.032E-15\\
  CAHA\_20320051\_402020 & 20:32:00.51 & +40:20:20.2 & 13.538 & 0.002 & 12.756 & 0.006 & 12.468 & 0.005 & 12.438 & 0.033 & 12.417 & 0.044 & 12.317 & 0.12 & 11.485 & 0.299 & --- & ---  & 1.365 & 5.296E-15\\
  CAHA\_20320956\_401901 & 20:32:09.56 & +40:19:01.3 & 18.248 & 0.009 & 13.633 & 0.006 & 11.067 & 0.016 & 9.439 & 0.017 & 8.995 & 0.017 & 8.7 & 0.039 & 8.721 & 0.137 & --- & ---  & 2.533 & 4.274E-15\\
  CAHA\_20320899\_401937 & 20:32:08.99 & +40:19:37.3 & 13.226 & 0.003 & 12.039 & 0.021 & 11.586 & 0.016 & 11.278 & 0.021 & 11.204 & 0.029 & 11.309 & 0.203 & 12.854 & 3.984 & --- & ---  & 1.803 & 3.219E-14\\
  CAHA\_20321374\_401856 & 20:32:13.74 & +40:18:57.0 & 18.195 & 0.007 & 14.942 & 0.007 & 13.283 & 0.005 & 12.483 & 0.081 & 12.185 & 0.058 & 11.035 & 0.227 & --- & ---  & --- & ---  & 2.548 & 1.6081E-14\\
  CAHA\_20322030\_401901 & 20:32:20.30 & +40:19:01.8 & 15.208 & 0.003 & 13.298 & 0.005 & 12.42 & 0.005 & 11.744 & 0.034 & 11.482 & 0.027 & 11.051 & 0.239 & --- & ---  & --- & ---  & 2.621 & 8.350E-15\\
  CAHA\_20322943\_401858 & 20:32:29.43 & +40:18:58.9 & 15.257 & 0.011 & 13.795 & 0.007 & 13.134 & 0.006 & 12.803 & 0.023 & 12.699 & 0.025 & 12.67 & 0.107 & 12.967 & 0.732 & --- & ---  & 1.584 & 2.927E-15\\
  CAHA\_20322803\_401927 & 20:32:28.03 & +40:19:27.6 & 15.742 & 0.012 & 13.726 & 0.017 & 12.758 & 0.024 & 12.166 & 0.021 & 11.895 & 0.042 & 11.74 & 0.115 & --- & ---  & --- & ---  & 1.584 & 4.456E-15\\
  CAHA\_20321720\_401910 & 20:32:17.20 & +40:19:10.0 & 15.663 & 0.003 & 15.001 & 0.006 & 14.695 & 0.006 & 14.39 & 0.175 & 14.528 & 0.198 & --- & ---  & --- & ---  & --- & ---  & 1.584 & 3.054E-15\\
  CAHA\_20324072\_402146 & 20:32:40.72 & +40:21:46.2 & 14.345 & 0.003 & 12.705 & 0.006 & 12.007 & 0.022 & 11.619 & 0.016 & 11.456 & 0.016 & 11.386 & 0.027 & 11.445 & 0.059 & --- & ---  & 2.227 & 1.829E-14\\
  CAHA\_20323427\_401851 & 20:32:34.27 & +40:18:51.1 & 15.769 & 0.005 & 13.771 & 0.005 & 12.887 & 0.006 & 12.478 & 0.017 & 12.353 & 0.019 & 12.112 & 0.08 & 12.208 & 0.269 & --- & ---  & 1.716 & 1.399E-15\\
  CAHA\_20320116\_402233 & 20:32:01.16 & +40:22:33.9 & 15.79 & 0.002 & 14.135 & 0.008 & 13.341 & 0.006 & 12.809 & 0.02 & 12.754 & 0.021 & 12.698 & 0.059 & 13.085 & 0.224 & --- & ---  & 1.978 & 3.314E-15\\
  TWOM\_20324184\_4014001 & 20:32:41.84 & +40:14:00.2 & 11.113 & 0.021 & 10.838 & 0.018 & 10.789 & 0.017 & 10.793 & 0.019 & 10.816 & 0.021 & 10.639 & 0.116 & 10.248 & 0.302 & --- & ---  & 0.883 & 4.458E-15\\
  TWOM\_20330733\_4014348 & 20:33:07.33 & +40:14:34.8 & 11.646 & 0.022 & 11.18 & 0.019 & 11.041 & 0.018 & 10.981 & 0.016 & 10.968 & 0.016 & 10.935 & 0.021 & 10.923 & 0.044 & --- & ---  & 1.131 & 2.840E-14\\
  TWOM\_20323631\_4020142 & 20:32:36.32 & +40:20:14.3 & 11.918 & 9.999 & 11.508 & 0.027 & 11.443 & 0.026 & 11.432 & 0.016 & 11.48 & 0.017 & 11.409 & 0.073 & 11.268 & 0.231 & --- & ---  & 0.986 & 3.822E-15\\
  TWOM\_20330655\_4022485 & 20:33:06.55 & +40:22:48.6 & 10.945 & 0.022 & 10.767 & 0.019 & 10.719 & 0.018 & 10.708 & 0.016 & 10.721 & 0.016 & 10.73 & 0.019 & 10.804 & 0.04 & --- & ---  & 0.898 & 3.0744E-15\\
  CAHA\_20321512\_401714 & 20:32:15.12 & +40:17:14.8 & 14.779 & 0.003 & 13.287 & 0.005 & 12.676 & 0.008 & 12.291 & 0.019 & 12.159 & 0.022 & 12.094 & 0.135 & --- & ---  & --- & ---  & 1.789 & 5.472E-15\\
  CAHA\_20320777\_401759 & 20:32:07.77 & +40:17:59.4 & 16.206 & 0.092 & 13.979 & 0.038 & 13.316 & 0.009 & --- & ---  & --- & ---  & --- & ---  & --- & ---  & --- & ---  & 4.417 & 2.468E-15\\
  CAHA\_20323293\_401511 & 20:32:32.93 & +40:15:11.6 & 17.384 & 0.011 & 15.979 & 0.01 & 15.275 & 0.011 & --- & ---  & --- & ---  & --- & ---  & --- & ---  & --- & ---  & 2.650 & 1.288E-13\\
  CAHA\_20322779\_401512 & 20:32:27.79 & +40:15:12.4 & 18.761 & 0.011 & 16.498 & 0.009 & 15.323 & 0.011 & --- & ---  & --- & ---  & --- & ---  & --- & ---  & --- & ---  & 5.351 & 4.285E-15\\
  CAHA\_20330360\_401707 & 20:33:03.60 & +40:17:07.4 & 20.784 & 0.065 & 19.02 & 0.044 & 18.445 & 0.094 & --- & ---  & --- & ---  & --- & ---  & --- & ---  & --- & ---  & 2.562 & 7.871E-15\\
  CAHA\_20315779\_401717 & 20:31:57.79 & +40:17:17.9 & 15.757 & 0.004 & 14.392 & 0.006 & 13.699 & 0.015 & --- & ---  & --- & ---  & --- & ---  & --- & ---  & --- & ---  & 1.920 & 3.300E-15\\
  CAHA\_20324085\_401304 & 20:32:40.85 & +40:13:04.7 & 20.26 & 0.069 & 17.919 & 0.041 & 17.023 & 0.049 & --- & ---  & --- & ---  & --- & ---  & --- & ---  & --- & ---  & 1.949 & 2.980E-15\\

 \enddata
\label{tab:ClassIII}
\end{deluxetable}

\end{landscape}
%\clearpage
%% To help institutions obtain information on the effectiveness of their
%% telescopes, the AAS Journals has created a group of keywords for telescope
%% facilities. A common set of keywords will make these types of searches
%% significantly easier and more accurate. In addition, they will also be
%% useful in linking papers together which utilize the same telescopes
%% within the framework of the National Virtual Observatory.
%% See the AASTeX Web site at http://aastex.aas.org/
%% for information on obtaining the facility keywords.

%% After the acknowledgments section, use the following syntax and the
%% \facility{} macro to list the keywords of facilities used in the research
%% for the paper.  Each keyword will be checked against the master list during
%% copy editing.  Individual instruments or configurations can be provided 
%% in parentheses, after the keyword, but they will not be verified.

%% Appendix material should be preceded with a single \appendix command.
%% There should be a \section command for each appendix. Mark appendix
%% subsections with the same markup you use in the main body of the paper.

%% Each Appendix (indicated with \section) will be lettered A, B, C, etc.
%% The equation counter will reset when it encounters the \appendix
%% command and will number appendix equations (A1), (A2), etc.

\end{document}